\documentclass{ws1-procs9x6}
\def\ga{\mathrel{\raise.3ex\hbox{$>$\kern-.75em\lower1ex\hbox{$\sim$}}}}
\def\la{\mathrel{\raise.3ex\hbox{$<$\kern-.75em\lower1ex\hbox{$\sim$}}}}
\def\beq{\begin{equation}}
\def\eeq{\end{equation}}
\def\he#1{\hbox{${}^{#1}$He}}
\def\li#1{\hbox{${}^{#1}$Li}}
\def\kmsmpc{{\,\rm km\,s^{-1}Mpc^{-1}}}
\newcommand\EE[2]{{#1}\!\times\! 10^{#2}}
\def\PL{{\it Phys. Lett.} }
\def\PR{{\it Phys. Rev.} }
\def\PRL{{\it Phys. Rev. Lett.} }
\def\NP{{\it Nucl. Phys.} }
\usepackage{epsf}
\usepackage{epsfig}
\begin{document}

\title{TASI Lectures on Dark Matter\footnote{\uppercase{S}ummary of lectures given   
at the \uppercase{T}heoretical \uppercase{A}dvanced \uppercase{S}tudy \uppercase{I}nstitute in \uppercase{E}lementary 
\uppercase{P}article \uppercase{P}hysics at the \uppercase{U}niversity of \uppercase{C}olorado at \uppercase{B}oulder - \uppercase{J}une 2-28,  
2002.}}

\author{Keith A. Olive\footnote{ \uppercase{T}his work was supported in part by 
\uppercase{DOE} grant \uppercase{DE-FG02-94ER40823} at \uppercase{M}innesota.}}

\address{William I. Fine Theoretical Physics
Institute, School of Physics and Astronomy, \\
University of Minnesota, Minneapolis, MN 55455 USA \\ 
E-mail: olive@umn.edu}


\maketitle

\abstracts{
\vskip -2.5in
\rightline{astro-ph/0301505}
\rightline{UMN--TH--2127/03}
\rightline{TPI--MINN--03/02}
\rightline{January 2003}
\vskip 1.9in
Observational evidence and theoretical motivation for dark matter
are presented and connections to the CMB and BBN are made.
Problems for baryonic and neutrino dark matter are summarized.
Emphasis is placed on the prospects for supersymmetric dark matter.}

\section{Lecture 1}

The nature and identity of the dark matter of the Universe is one of the
most challenging problems facing modern cosmology. The problem is a
long-standing one, going back to early observations of mass-to-light
ratios by Zwicky\cite{zw}. 
Given the distribution (by number) of galaxies with total 
luminosity $L$,  $\phi(L)$, one can 
compute the mean luminosity density
of galaxies
\beq
	{\mathcal L}  = \int L \phi(L) dL	
\eeq
which is determined to be\cite{meanl}
\beq
	  {\mathcal L} \simeq 2 \pm 0.2 \times 10^8  h_o L_\odot Mpc^{-3} 	
\eeq
where  $L_\odot = 3.8 \times 10^{33}$   erg s$^{-1}$ is the solar 
luminosity.  
In the absence of a cosmological constant, one can 
define a critical energy density, $	\rho_c  = 3H^2 / 8 \pi G_N	=  1.88
\times 10^{-29} {h_o}^2$ g cm$^{-3}$,
such that $\rho =\rho_c$  for three-space curvature $k = 0$, 
where the present value of the Hubble 
parameter has been defined by $H_o = 100 h_o$ km Mpc$^{-1}$ s$^{-1}$.	
We can now define a critical mass-to-light ratio is given by
\beq
	(M/L)_c = \rho_c/{\mathcal L}  \simeq  1390 h_o (M_\odot/L_\odot)	
\eeq
which can be used to determine the cosmological density parameter
\beq
	\Omega_m =  {\rho \over
\rho_c} = (M/L)/(M/L)_c 	
\eeq

Mass-to-light ratios are, however, strongly dependent on the distance
scale on which they are determined\cite{fg}. In the solar neighborhood
$M/L \simeq 2 \pm 1$
 (in solar units), yielding values of $\Omega_m$  of only $\sim .001$.  
In the bright central parts of galaxies, $M/L  \simeq  (10-20)h_o$ 
so that $\Omega_m  \sim 0.01$.  On larger scales, that of
 binaries and small groups of galaxies, $M/L \sim (60-180)h_o$ 
and $\Omega_m   \simeq  0.1$.  On even larger scales, that of clusters,
$M/L$ may  be as large as $(200-500)h_o$ giving $\Omega_m \simeq 0.3$. 
 This progression in $M/L$ seems to have halted, as
even on the largest scales observed today, mass-to-light ratios imply
values of $\Omega_m \la 0.3 - 0.4$.  Thus when one considers the scale of
galaxies (and their halos) and larger, the presence of dark matter (and
as we shall see, non-baryonic dark matter) is required.

Direct observational
evidence for dark matter is found from a variety of sources. On the scale
of galactic halos, the  observed flatness of the rotation curves of
spiral galaxies is a clear indicator for dark matter. There is also
evidence for dark matter in elliptical galaxies, as well as clusters of
galaxies coming from the X-ray observations of these objects. Also, direct
evidence has been obtained through the study of gravitational lenses. On
the theoretical side, we  predict the presence of dark matter (or dark
energy) because 1) it is a strong prediction of most inflation models (and
there is at present no good alternative to inflation)
 and 2) our current understanding of galaxy 
formation requires substantial amounts of dark matter to account for
the growth of density fluctuations.  
	One can also make a strong case for the existence of non-baryonic 
dark matter in
particular. The recurrent problem  with baryonic dark matter
is that not only is it very difficult to hide baryons, but 
given the amount of dark matter required on large scales, there is a
direct conflict with primordial
nucleosynthesis if all of the dark matter is baryonic. 
 In this first lecture, I will review the observational and theoretical
evidence supporting the existence of dark matter.

\subsection{Observational Evidence}

\begin{figure}[t]
\hskip 2cm
\epsfxsize=6cm   %width of figure - will enlarge/reduce the figures
\epsfbox{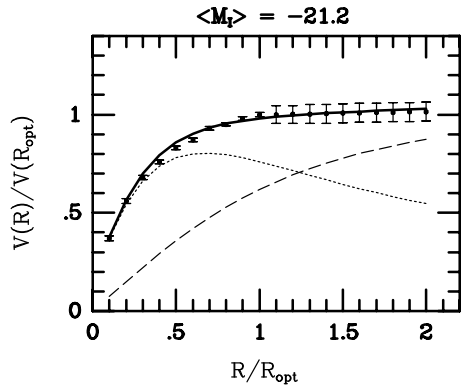}
%\figurebox{2cm}{3cm}{} %to have a box alone 
%\centerline{\epsfxsize=3.9in\epsfbox{procs-fig1.eps}}   
\caption{Synthetic rotation curve\protect\cite{pss} for galaxies with
$\langle M
\rangle = -21.2$.  The dotted curve shows the disk contribution,
whereas the dashed curve shows the halo contribution.  
\label{rot}}
\end{figure}

 Assuming that galaxies are in virial equilibrium,
 one expects that one can relate 
the mass at a given distance $r$, from the center of 
a galaxy to its rotational velocity by
\beq
	M(r) \propto v^2 r/G_N 	
\eeq
The rotational velocity, $v$, is measured\cite{fg,rft}
 by observing 21 cm 
emission lines in HI regions (neutral hydrogen) beyond the point 
where most of the light in the galaxy ceases.  A subset of a
compilation\cite{pss}  of nearly 1000
 rotation curves of spiral galaxies is shown in Fig. \ref{rot}. The
subset shown is restricted to a narrow range in brightness, but is
characteristic for a wide range of spiral galaxies. Shown is the
rotational velocity as a function of
$r$ in units of the optical radius.  If the bulk of the mass is 
associated with light, then beyond the point where most of the light 
stops, $M$ would be  constant and $v^2  \propto 1/r$.  This is not the
case, as the rotation
 curves appear to be flat, i.e., $v \sim$ constant outside the
 core of the galaxy. This implies that $M \propto r$ beyond the point
 where the light stops.  This is one of the strongest pieces of 
evidence for the existence of dark matter. Velocity measurements indicate
dark matter in elliptical galaxies as well\cite{sag}.

Galactic rotation curves are not the only observational indication for the
existence of dark matter.  X-ray emitting hot gas in elliptical galaxies 
also provides an important piece of evidence for dark matter. 
 A particularly striking example is that of the large elliptical M87. 
Detailed profiles of the temperature and density of the hot X-ray
emitting gas have been mapped out\cite{fgo}. Assuming hydrostatic
equilibrium, these measurements allow  one to determine 
the overall mass distribution in the galaxy necessary to bind the hot gas.
Based on an isothermal model with temperature $kT = 3$keV (which leads
to a conservative estimate of the total mass), Fabricant and
 Gorenstein\cite{fgo}
predicted that the total mass out to a radial distance
 of 392 kpc is $5.7 \times 10^{13} M_\odot$,
whereas the mass in the hot gas is only $2.8 \times 10^{12} 
M_\odot$ or only 5\%
of the total. The visible mass is expected to contribute 
only 1\% of the total.

M87 is not the only example of an elliptical galaxy in which  X-ray
emitting hot gas is observed to indicate the presence of dark matter.
X-ray observations have shown that the total mass associated with
elliptical galaxies is considerably larger than the luminous component
in many examples of varying morphological types\cite{ell}. 
Mass-to-light ratios for these systems vary, with most being
larger than 30$h_0$ and some ranging as high as $\sim 200 h_0$. 
  
In addition,  similar inferences pertaining to the existence of dark
matter can be made from the X-ray emission  from small groups of
galaxies\cite{mush}. On these scales, mass-to-light ratios are
typically $\ga 100 h_0$ and detailed studies have shown that the
baryon faction in these systems is rather small. Furthermore, it was
argued\cite{white} that cluster baryon fractions should not differ from
the Universal value given by $\Omega_B/\Omega_m$. Using an estimate of
$\Omega_B = 0.04$ from big bang nucleosynthesis (BBN, see below), and
baryon fractions ranging from 0.1 to 0.3, one would obtain an estimate
for the total matter density of 0.13 -- 0.4.

Another piece of evidence on large scales is available from 
gravitational lensing\cite{tyson}. 
The systematic lensing of the roughly 150,000 
galaxies per deg$^2$
at redshifts between $z = 1 - 3$ into arcs and arclets allow 
one to trace the
matter distribution in a foreground cluster. Lensing observations 
can be categorized as either strong (multiple images) or weak
(single images).  Both require the presence of a dominant dark matter
component. 

Strong lensing is particularly adept in testing the overall geometry of
the Universe\cite{turner,glose}.  While a cluster which provides
multiple lenses of a single background galaxy (at known redshift) is
useful for determining the total cluster mass, when several background
galaxies are lensed, it is possible to constrain the values of $\Omega_m$
and $\Omega_\Lambda$ \cite{blan}. The recent results of
\cite{glose} show a degeneracy in the  $\Omega_m$
-- $\Omega_\Lambda$ plane. Nevertheless,  the allowed region is offset
from similar types of degeneracies found in supernovae searches and the
CMB (see below). Indeed, these lensing
results are much more constraining in
$\Omega_m$ than the other techniques, though a residual uncertainty of
about 30 \% persists.  While in principle, these results find that any
value (from 0 to 1) is possible for $\Omega_\Lambda$, $\Omega_m < 0.5$ for
low values of $\Omega_\Lambda$ and $\Omega_m < 0.4$ for
higher values of $\Omega_\Lambda$ ($\ga 0.6$).

Weak lensing of galaxies by galaxies can (on a statistical basis)
also probe the nature of galactic halos.  Recent studies based on weak
lensing data indicate that galactic halos may be far more  extended than
previously thought\cite{weakgg} (radii larger than 200 $h_0^{-1}$ kpc). 
These results also imply a substantial contribution to 
$\Omega_m$ (of order 0.1-0.2) on this scale.  On larger scales,
using many cluster lenses enables one to estimate $\Omega \simeq 0.3$
\cite{mell}. Another use of weak lensing statistics is to determine
the evolution of cosmic shear and hence an estimate of $\Omega_m$
\cite{vanw}.   Finally, there exist a number of examples of dark
clusters, ie., lenses with no observable counterpart\cite{dark}.
The contribution of these objects (if they are robust) to $\Omega_m$ is
not clear.  For a recent review of weak lensing see \cite{mell2}.

Finally, on very large scales, it is possible to get an estimate of
$\Omega_m$ from the distribution of peculiar velocities of galaxies and
clusters. On scales,
$\lambda$,  where perturbations, $\delta$, are still
small, peculiar velocities can be expressed\cite{peeb}
 as $v \sim H \lambda \delta \Omega_m^{0.6}$.
On these scales, older measurements of the peculiar velocity field from 
the IRAS galaxy catalogue indicate that indeed $\Omega$ 
is close to unity\cite{iras}. Some of the new data
indicates a lower value in the range\cite{newiras}  0.2 -- 0.5, but does 
not conclusively exclude $\Omega_m = 1$ \cite{morenew}.

The above discussion of observational evidence for dark matter pertains
largely to the overall matter density of the Universe $\Omega_m$.
(For a comprehensive review of determinations of the matter density, see
ref. \cite{rev}. )  However, the matter density and the overall {\em
curvature} are not related one-to-one.
The expansion rate of the Universe in the standard FRW model
is expressed by the Friedmann equation
\beq
H^2 = {{\dot R}^2 \over R^2} = {8 \pi G_N \rho \over 3} - {k \over R^2} +
{\Lambda \over 3}
\label{fried}
\eeq  
where $R(t)$ is the cosmological scale factor, $k$ is the three-space
curvature constant ($k = 0, +1, -1$ for a spatially flat, closed or open
Universe), and $\Lambda$ is the cosmological constant.         
The Friedmann equation can be rewritten as
\beq
	(\Omega - 1)H^2  = {k \over R^2}	
\label{o-1}	
\eeq
so that $k = 0, +1, -1$ corresponds to $\Omega = 1, \Omega > 1$
 and $\Omega < 1$.  However, the value of $\Omega$ appearing in Eq.
(\ref{o-1}) represents the sum $\Omega = \Omega_m + \Omega_\Lambda$
of contributions from the matter density ($\Omega_m$) and the
cosmological constant $(\Omega_\Lambda = \Lambda/3H^2)$.

There has been a great deal of progress in the last several years
concerning the determination of both $\Omega_m$ and $\Omega_\Lambda$.
Cosmic Microwave Background (CMB) anisotropy experiments have been able
to determine the curvature (i.e. the sum of $\Omega_m$ and
$\Omega_\Lambda$) to with in about 10\%, while observations of type Ia
supernovae at high redshift provide information on a (nearly) orthogonal
combination of the two density parameters.

The CMB is of course deeply rooted in the development and verification of
the big bang model\cite{ah}.  Indeed, it was the formulation of BBN that
led to the prediction of the microwave background.  The argument is rather
simple. BBN requires temperatures greater than 100 keV, which according
to the standard model time-temperature relation,
$t_{\rm s} T^2_{\rm MeV} = 2.4/\sqrt{N}$, where
$N$ is the number of relativistic degrees of freedom at temperature
$T$, and corresponds to timescales less than about 200 s. The typical
cross section for the first link in the nucleosynthetic chain is
\beq 
\sigma v (p + n \rightarrow D + \gamma) \simeq 5 \times 10^{-20} 
{\rm cm}^3/{\rm s}
\eeq
This implies that it was necessary to achieve a density
\beq
n \sim {1 \over \sigma v t} \sim 10^{17} {\rm cm}^{-3}
\eeq
for nucleosynthesis to begin.
The density in baryons today is known approximately from the density of
visible matter to be ${n_B}_o \sim 10^{-7}$ cm$^{-3}$ and since
we know that that the density $n$ scales as $R^{-3} \sim T^3$, 
the temperature today must be
\beq
T_o = ({n_B}_o/n)^{1/3} T_{\rm BBN} \sim 10 {\rm K}
\eeq
thus linking two of the most important tests of the Big Bang theory.

\begin{figure}[htbp]
\hspace{0.5truecm}
\centering
\epsfxsize=10cm   %width of figure - will enlarge/reduce the figures
\epsfbox{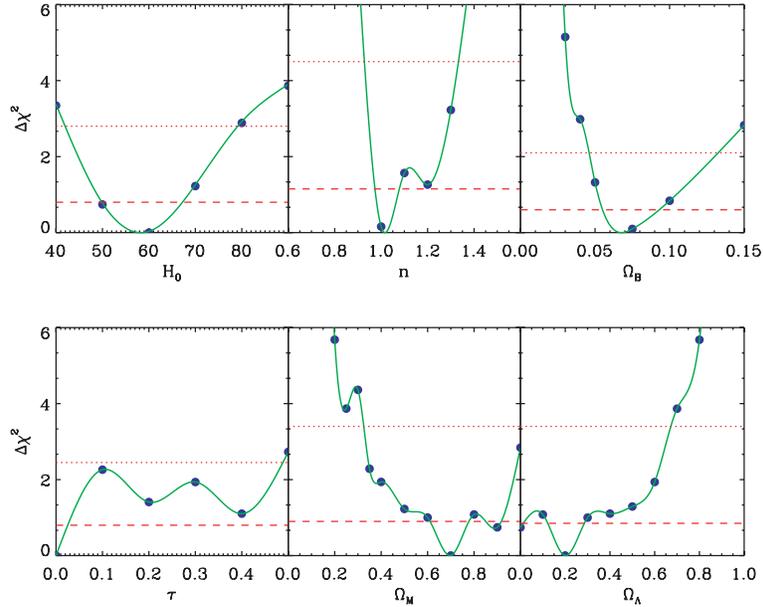}
\caption{{$\Delta\chi^2$ calculated 
with the MAXIMA-1 and COBE data as a function of parameter value.  Solid
blue circles show grid points in parameter space, and the green lines
were obtained by interpolating between grid points.  The parameter values
where the green line intercepts the red dashed (dotted) line corresponds
to the 68\% (95\%) frequentist confidence region\protect\cite{freq}. }}
\label{freq1}
\end{figure}

Microwave background anisotropy measurements have made tremendous
advances in the last few years. The power spectrum\cite{max,boom,dasi1}
has been measured relatively accurately out to multipole moments
corresponding to $\ell \sim 1000$. The details of this spectrum enable
one to make accurate predictions of a large number of fundamental
cosmological parameters\cite{boom,max2,dasi2,cbi,vsa,arch}. An example of
these results as found by a recent frequentist analysis\cite{freq} is
shown in Fig.~\ref{freq1}.

Of particular interest to us here is the CMB determination of the total
density, $\Omega_{\rm tot}$, as well as the matter density $\Omega_m$. 
The results of recent CMB anisotropy measurements are summarized in Table
1. As one can see, there is strong evidence that the Universe is flat or
very close to it. Furthermore, the matter density is very consistent with
the observational determinations discussed above and the baryon density,
as we will see below, is consistent with the BBN production of D/H and its
abundance in quasar absorption systems. 
\begin{table}[h]
\tbl{Results from recent CMB anisotropy measurements.\vspace*{1pt}}
{\footnotesize
%\tabcolsep7pt
%\begin{tabular}{@{}crrrr@{}}
\begin{tabular}{|c|c|c|c|}
\hline
{} &{} &{} &{} \\[-1.5ex]
{} & $\Omega_{\rm tot}$ & $\Omega_m h^2$ & $\Omega_B h^2$ \\[1ex]
\hline
{} &{} &{} &{} \\[-1.5ex]
BOOMERanG\cite{boom} &$1.03 \pm 0.06$ &$0.12 \pm 0.05$
&$0.021^{+0.004}_{-0.003}$ \\[1ex]
MAXIMA\cite{max2} &$0.9^{+0.09}_{-0.08}$ &$0.17^{+0.08}_{-0.04}$
&$0.0325\pm 0.0125$ \\[1ex] 
MAXIMA (freq.)\cite{freq} &$0.89^{+0.13}_{-0.10}$ &$0.25^{+0.07}_{-0.09}$
&$0.026^{+0.010}_{-0.006}$\\[1ex]
DASI\cite{dasi2} &$1.04 \pm 0.06$ &$0.14 \pm 0.04$
&$0.022^{+0.004}_{-0.003}$ \\[1ex]
CBI\cite{cbi} &$0.99 \pm 0.12$ &$0.17^{+0.08}_{-0.06}$
&$0.022^{+0.015}_{-0.009}$ \\[1ex]
VSA\cite{vsa} &$1.03 \pm 0.012$ &$0.13^{+0.08}_{-0.05}$ &$0.029 \pm 0.009$
\\[1ex]
Archeops\cite{arch} & $1.16^{+0.24}_{-0.20}$ & -- &$
0.019^{+0.006}_{-0.007}$\\[1ex]
\hline
\end{tabular}\label{tab1} }
%\vspace*{-13pt}
\end{table}

\begin{figure}
\hskip 1.5cm
\epsfxsize=8cm   %width of figure - will enlarge/reduce the figures
\epsfbox{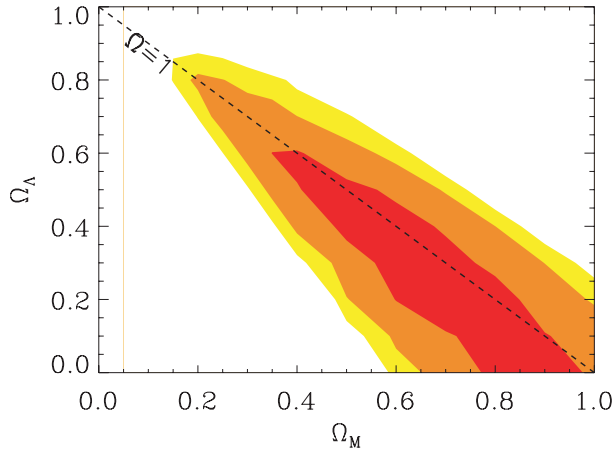}
\caption{\footnotesize Two-dimensional frequentist confidence regions
in the ($\Omega_{\rm M},\Omega_{\Lambda}$) plane.  The red, orange and yellow 
regions correspond to the 68\%, 95\%, and 99\% confidence
regions respectively.  The dashed black line corresponds
to a flat universe, $\Omega=\Omega_m + \Omega_{\Lambda}=1$.}
\label{lm}
\end{figure}

The discrepancy between the CMB value of $\Omega_m$ and
$\Omega_B$ is sign that non-baryonic matter (dark matter) is
required.  Furthermore, the apparent discrepancy between the CMB value
of $\Omega_{\rm tot}$ and $\Omega_m$, though not conclusive on its own,
is a sign that a contribution from the vacuum energy density or
cosmological constant, is also required.  The preferred region in the
$\Omega_m - \Omega_\Lambda$ plane as determined by a frequentist analysis
of MAXIMA data is shown in Fig. \ref{lm}.

The presence or absence of a cosmological constant is a long standing
problem in cosmology.  To theorists, it is particularly offensive
due to the necessary smallness of the constant. We know that 
the cosmological term is at most a factor of a few times larger than the
current mass density.  Thus from Eq. (\ref{fried}), we see that the
dimensionless combination, $G_N \Lambda \la 10^{-121}$. Nevertheless,
even a small non-zero value for $\Lambda$ could greatly affect the
future history of the Universe: allowing open Universes to recollapse (if
$\Lambda < 0 $), or closed Universes to expand forever (if $\Lambda > 0$
and sufficiently large). 

\begin{figure}
\hskip 1.5cm
\epsfxsize=8cm   %width of figure - will enlarge/reduce the figures
\epsfbox{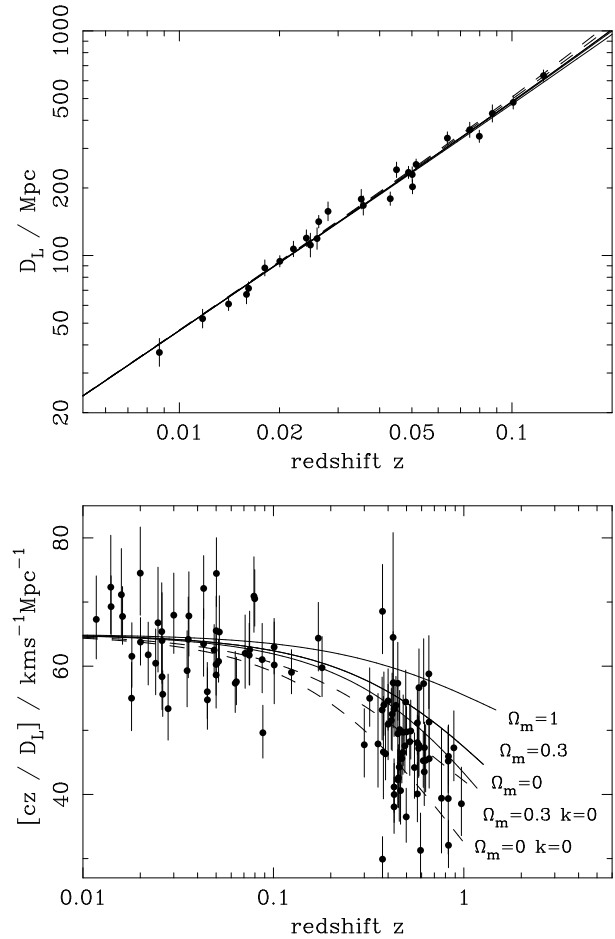}
\caption{The type Ia supernova Hubble diagram\protect\cite{sn1}
taken from \protect\cite{op}. The first panel shows that, for $z\ll 1$,
the large-scale Hubble flow is indeed linear and uniform;
the second panel shows an expanded scale, with the linear trend
divided out, and with the redshift range extended to show
how the Hubble law becomes nonlinear. 
Comparison with the prediction of Friedmann-Lema\^itre models appears to
favor  a vacuum-dominated universe.}
\label{snhub}
\end{figure}

Another exciting development has been
the use of type Ia supernovae,  which now allow
measurement of relative distances with 5\% precision.
In combination with Cepheid data from the HST key
project on the distance scale, SNe results are the dominant contributor
to the best modern value for $H_0$: $72\kmsmpc \pm 10\%$ \cite{freedh0}.
Better still, the analysis of
high-$z$ SNe has allowed the first meaningful
test of cosmological geometry to be carried out, as shown in Fig.
\ref{snhub}.

\begin{figure}
\hskip 2cm
\epsfxsize=8cm   %width of figure - will enlarge/reduce the figures
\epsfbox{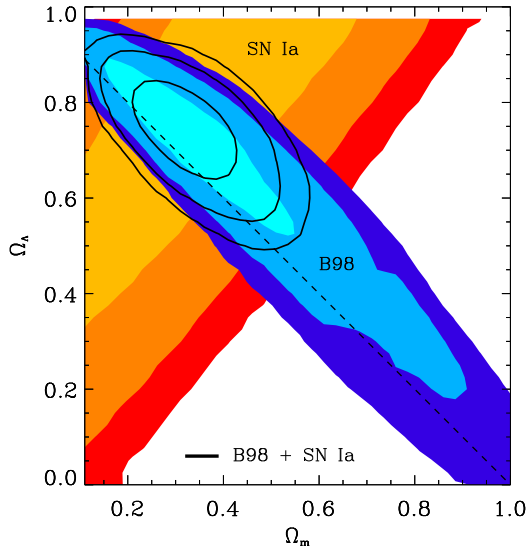}
\caption{Likelihood-based
confidence contours\protect\cite{boom} over the plane $\Omega_\Lambda$
(i.e. $\Omega_v$ assuming $w=-1$ vs $\Omega_m$.
The SNe Ia results very nearly constrain $\Omega_v-\Omega_m$,
whereas the results of CMB anisotropies (from the Boomerang 98 data) favor
a flat model with $\Omega_v+\Omega_m\simeq 1$. The intersection of these
constraints is the most direct (but far from the only) piece
of evidence favoring a flat model with $\Omega_m\simeq 0.3$.
\label{conf}
}
\end{figure}

These results can be contrasted with those from the CMB anisotropy
measurements as in Fig. \ref{conf}.  We are led to a seemingly conclusive
picture.  The Universe is nearly flat with $\Omega_{\rm tot} \simeq 1$.
However the density in matter makes up only 20-50\% of this total, with
the remainder in a cosmological constant or some other form of dark
energy.

\subsection{Theory}

Theoretically, there is no lack of support for the dark matter hypothesis.
The standard big bang model including inflation almost requires 
$\Omega_{\rm tot} = 1~$\cite{infl}. This can be seen from the following
simple solution to the curvature problem.
 The simple and
unfortunate fact that at present we do not even know whether $\Omega$ is
larger or smaller than one, indicates that we do not know the sign of
the curvature term further implying that it is subdominant in Eq.
(\ref{fried})
\beq
	  { k \over R^2 } < {8 \pi G \over 3} \rho
\eeq
In an adiabatically expanding Universe, $R \sim T^{-1}$   where $T$ is the
temperature of the thermal photon background.  Therefore the quantity
\beq
	\hat{k} = { k \over R^2 T^2} <  {8 \pi G \over 3 T_o^2} < 2 \times 10^{-58}
\label{khat}
\eeq
is dimensionless and constant in the standard model.  This is known as
the curvature problem and can be resolved by a period of inflation.
Before inflation, let us write $R = R_i$, $T = T_i$  and $R \sim T^{-1}$.  During
inflation, $R \sim T^{-1} \sim e^{Ht}$, where $H$ is constant.  After inflation, $R =
R_f  \gg R_i$  but $T = T_f  = T_R  \la T_i$  where $T_R$  is the temperature
 to which the
Universe reheats.  Thus $R \not\sim T$   and $\hat{k} \rightarrow 0$
 is not constant.  But from
Eqs. (\ref{o-1}) and (\ref{khat}) if $\hat{k} \rightarrow 0$ then
 $\Omega \rightarrow 1$, and since typical
inflationary models contain much more expansion than is necessary, $\Omega$
becomes exponentially close to one.

	The inflationary prediction of $\Omega = 1$ is remarkably consistent
with the CMB measurements discussed above. Furthermore,  we know two
things:  Dark matter exists, since we don't see
$\Omega = 1$ in luminous objects, and most (about 90\%) of the dark
matter is not baryonic.  The latter conclusion is a result of our
forthcoming discussion on BBN which constrains the baryon-to-photon ratio
and hence
$\Omega_B$.  Thus $1-\Omega_B$ is not only dark but
also non-baryonic. Furthermore, the matter density is surely composed
of several contributions: $\Omega_m = \Omega_B + \Omega_\nu + \Omega_\chi$
where the latter represents the dark matter contribution. 

	Another important piece of theoretical evidence for dark
 matter comes from the simple fact that we are  living in a galaxy.
The type of perturbations produced
 by inflation\cite{press} are, in most models,
 adiabatic perturbations ($\delta\rho/\rho \propto
 \delta T/T)$, and I
 will restrict my attention to these.  Indeed, the perturbations
produced by inflation
 also have the very nearly scale-free spectrum described by 
Harrison and Zeldovich\cite{hz}.  When produced, scale-free perturbations 
fall off as $\frac{\delta \rho}{\rho} \propto l^{-2}$
 (increase as the square of the wave number). 
 At early times $\delta\rho/\rho$ grows as $t$
 until the time when the horizon scale (which is
 proportional to the age of the Universe) is comparable to $l$.  At later 
times, the growth halts (the mass contained within the volume $l^3$  
has become smaller than the Jean's mass) and   
  $\frac{\delta \rho}{\rho} = \delta$ (roughly) independent of the scale $l$.
 When the Universe becomes matter dominated, the Jean's mass 
drops dramatically and growth continues as $\frac{\delta \rho}{\rho} \propto
 R \sim 1/T$. 

 The transition to matter 
dominance
is determined by setting the energy densities in radiation
(photons and any massless  neutrinos) equal to the energy density in  
matter (baryons and any dark matter).  For three massless  neutrinos and
baryons (no dark matter), matter dominance begins at
\beq
	T_m  = 0.22 m_B \eta	
\eeq
and for $\eta < 7 \times 10^{-10}$, this corresponds to
$T_m < 0.14$ eV.

	Because we are considering adiabatic perturbations,
 there will be anisotropies produced in the microwave 
background radiation on the order of $\delta T/T \sim \delta$.  
The value of $\delta$, the amplitude of the density fluctuations at horizon 
crossing, has been determined by COBE\cite{cobe}, $\delta =
(5.7 \pm 0.4) \times 10^{-6}$.  Without the existence of dark matter,
 $\delta \rho/\rho$ in baryons could then achieve a maximum value of only
$\delta\rho/\rho \sim A_\lambda \delta(T_m/T_o)  
\la 2 \times 10^{-3}A_\lambda$,
where $T_o = 2.35 \times 10^{-4}$ eV is the present temperature of the
microwave background and $A_\lambda
\sim 1-10$ is a scale dependent growth factor. 
 The overall growth in $\delta \rho / \rho$ is too small to argue
 that growth has entered a nonlinear regime needed to explain
 the large value ($10^5$) of $\delta\rho/\rho$ in galaxies.

	Dark matter easily remedies this dilemma in the following way.
 The transition to matter dominance is determined by setting equal 
to each other the energy densities in radiation (photons and any massless 
neutrinos) and matter (baryons and any dark matter). 
While if we suppose that there exists 
dark matter with an abundance $Y_\chi = n_\chi/n_\gamma$  
(the ratio of the number density of $\chi$'s to photons) then
\beq
	T_m  = 0.22 m_\chi Y_\chi	
\eeq
Since we can write $m_\chi Y_\chi/{\rm GeV} = \Omega_\chi
 h^2/(4 \times 10^7)$,
we have $T_m/T_o = 2.4 \times 10^4 \Omega_\chi h^2$ which is 
to be compared with
600 in the case of baryons alone.  
The baryons, although still bound to the radiation until 
decoupling,  now see deep potential wells formed by the dark matter
 perturbations to fall into and are no longer required to 
grow at the rate $\delta \rho/\rho \propto R$.

	With regard to dark matter and galaxy formation, all forms 
of dark matter are not equal.  They can be distinguished 
by their relative temperature at $T_m$ \cite{bond}. Particles which are still 
largely relativistic at $T_m$ (like neutrinos or other particles with 
$m_\chi < 100$ eV) have the property\cite{free} that 
(due to free streaming) they
erase perturbations 
 out to very large scales given by the Jean's mass
\beq
	M_J  = 3 \times 10^{18}  {M_\odot \over {m_\chi}^2(eV)}	
\label{mj}
\eeq
Thus, very large scale structures form first and galaxies
 are expected to fragment out later.  Particles with this 
property are termed hot dark matter particles.  
Cold particles ($m_\chi > 1$ MeV) have the opposite behavior. 
 Small scale structure forms first aggregating to form 
larger structures later. It is now well known that pure HDM cosmologies
can not reproduce the observed large scale structure of the Universe.
In contrast, CDM does much better. Current attention is focused on
so-called $\Lambda$CDM cosmologies based on the $\Omega_\Lambda-\Omega_m$
contribution to the curvature discussed above.

\section{Lecture 2}

\subsection{Big Bang Nucleosynthesis}

The standard model\cite{wssok} of big bang nucleosynthesis (BBN)
is based on the relatively simple idea of including an extended nuclear
network into a homogeneous and isotropic cosmology.  Apart from the
input nuclear cross sections, the theory contains only a single parameter,
namely the baryon-to-photon ratio,
$\eta$. Other factors, such as the uncertainties in reaction rates, and
the neutron mean-life can be treated by standard statistical and Monte
Carlo techniques\cite{kr,cfo}.  The theory then allows one to make
predictions (with well-defined uncertainties) of the abundances of the
light elements, D,
\he3, \he4, and \li7.

\subsubsection{Theory}
Conditions for the synthesis of the light elements were attained in the
early Universe at temperatures  $T \ga $ 1 MeV.  In the early Universe,
the energy density was dominated by radiation with
\beq
\rho = {\pi^2 \over 30} ( 2 + {7 \over 2} + {7 \over 4}N_\nu) T^4
\label{rho}
\eeq
from the contributions of photons, electrons and positrons, and $N_\nu$
neutrino flavors (at higher temperatures, other particle degrees of
freedom should be included as well). At these temperatures, weak
interaction rates were in equilibrium. In particular, the processes
\begin{eqnarray}
n + e^+ & \leftrightarrow  & p + {\bar \nu_e} \nonumber \\
n + \nu_e & \leftrightarrow  & p + e^- \nonumber \\
n  & \leftrightarrow  & p + e^- + {\bar \nu_e} 
\label{beta}
\end{eqnarray}
fix the ratio of
number densities of neutrons to protons. At $T \gg 1$ MeV, $(n/p) \simeq
1$. 

The weak interactions do not remain in equilibrium at lower temperatures.
Freeze-out occurs when the weak interaction rate, $\Gamma_{wk} \sim G_F^2
T^5$ falls below the expansion rate which is given by the Hubble
parameter, $H \sim \sqrt{G_N \rho} \sim T^2/M_P$, where $M_P =
1/\sqrt{G_N} \simeq 1.2
\times 10^{19}$ GeV. The $\beta$-interactions in eq.
(\ref{beta}) freeze-out at about 0.8 MeV.
 As the temperature falls
and approaches the point where the weak interaction rates are no longer
fast enough to maintain equilibrium, the neutron to proton ratio is given
approximately by the Boltzmann factor,
$(n/p)
\simeq e^{-\Delta m/T} \sim 1/6$, where $\Delta m$ is the neutron-proton
mass difference. After freeze-out, free neutron decays drop the ratio
slightly to about 1/7 before nucleosynthesis begins. 

The nucleosynthesis chain begins with the formation of deuterium
by the process, $p+n \rightarrow$ D $+~\gamma$.
However, because of the large number of photons relative to nucleons,
$\eta^{-1} = n_\gamma/n_B \sim 10^{10}$, deuterium production is delayed
past the point where the temperature has fallen below the deuterium
binding energy, $E_B = 2.2$ MeV (the average photon energy in a blackbody
is ${\bar E}_\gamma \simeq 2.7 T$).  This is because there are many
photons in the exponential tail of the photon energy distribution with
energies $E > E_B$ despite the fact that the temperature or ${\bar
E}_\gamma$ is less than $E_B$.  The degree to which deuterium production
is delayed can be found by comparing the qualitative expressions for the
deuterium production and destruction rates,
\begin{eqnarray}
\Gamma_p & \approx & n_B \sigma v \\ \nonumber
\Gamma_d & \approx & n_\gamma \sigma v e^{-E_B/T}
\end{eqnarray}
When the quantity $\eta^{-1}
{\rm exp}(-E_B/T)
\sim 1$, the rate for  deuterium destruction (D $+ ~\gamma
\rightarrow p + n$) finally falls below the deuterium
production rate and the nuclear chain begins at a temperature
$T \sim 0.1 MeV$.

The dominant product of big bang nucleosynthesis is \he4 and its
abundance is very sensitive to the
$(n/p)$ ratio
\beq
Y_p = {2(n/p) \over \left[ 1 + (n/p) \right]} \approx 0.25
\label{ynp}
\eeq
i.e., an
abundance of close to 25\% by mass. Lesser amounts of the other light
elements are produced: D and \he3 at the level of about $10^{-5}$ by
number,  and \li7 at the level of $10^{-10}$ by number. 

\begin{figure}[t]
%\hspace{0.2truecm}
\epsfxsize=12cm   %width of figure - will enlarge/reduce the figures
\epsfbox{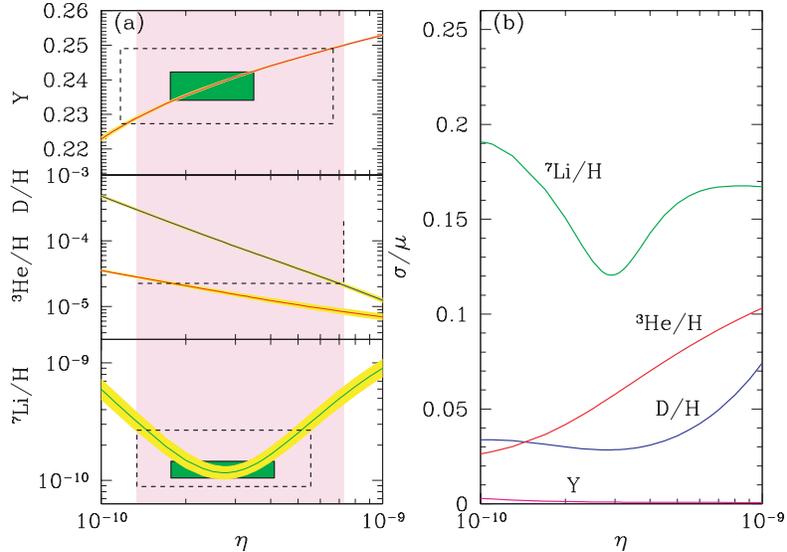}
\caption{{The light element abundances from big bang
nucleosynthesis as a function of $\eta_{10}$.}}
\label{nuc8}
\end{figure}

The resulting
abundances of the light elements\cite{cfo} are shown in Figure~\ref{nuc8},
over the range in
$\eta_{10} = 10^{10} \eta$ between 1 and 10.  The left plot shows the
abundance of \he4 by mass, $Y$, and the abundances of the other three
isotopes by number.  The curves indicate the central predictions from
BBN, while the bands correspond to the uncertainty in the predicted
abundances based primarily the uncertainty in the input  nuclear
reactions as computed by Monte Carlo in ref. \cite{cfo}.
This theoretical
uncertainty is shown explicitly in the right panel as a function of
$\eta_{10}$. The dark shaded boxes correspond to the observed
abundances  of \he4 and
\li7 and will be discussed below.  The dashed boxes correspond to the
ranges of the elements consistent with the systematic uncertainties in
the observations. The broad band shows a liberal range for $\eta_{10}$
consistent with the observations. At present, there is a general concordance between the theoretical
predictions and the observational data.

\vskip .5in

\subsubsection{Abundances}

In addition to it BBN production, \he4~is made in stars, and thus
co-produced with heavy elements. Hence the best sites for determining the
primordial
\he4 abundance are in metal-poor regions of hot, ionized gas in nearby
external galaxies (extragalactic H{\small II} regions). Helium indeed
shows a linear correlation with metallicity in these systems, and the
extrapolation to zero metallicity gives the primordial abundance
(baryonic mass fraction)\cite{fo98}
\beq
Y_p = 0.238 \pm 0.002 \pm 0.005.
\label{he4}
\eeq
Here, the first error is statistical and reflects the large sample of
systems, whilst the second error is systematic and dominates.

The systematic uncertainties in these observations have not been
thoroughly explored to date\cite{OSk}. In particular, there may be reason
to suspect that the above primordial abundance will be increased due to
effects such as underlying stellar absorption in the H{\small II} 
regions. We
note that other analyses give similar results: $Y_p = 0.244 \pm 0.002 \pm
0.005$~\cite{izotov} and 0.239 $\pm 0.002$~\cite{peim}.  

The primordial \li7 abundance comes from measurements in the atmospheres
of primitive (Population II) stars in the stellar halo of our Galaxy. The
\li7/H abundance is found to be constant for stars with low metallicity,
indicating a primordial component, and a recent determination gives
\beq
\label{rfbon}
\frac{\li7}{\rm H}_p = \EE{(1.23 \pm 0.06_{-0.32}^{+0.68})}{-10}\ ({\rm
95\% \ CL}),
\label{firstLi}
\eeq
where the small statistical error is overshadowed by systematic
uncertainties\cite{ryan}.  The range (\ref{rfbon}) may, however, be
underestimated, as a recent determination\cite{liglob} uses a different
procedure to determine stellar atmosphere parameters, and gives
${\li7/{\rm H}}_p = (2.19 \pm 0.28) \times 10^{-10}$. At this stage, it
is not possible to determine which method of analysis is more accurate,
indicating the likelihood that the upper systematic uncertainty
in (\ref{rfbon}) has been underestimated.  Thus, in order to
obtain a conservative bound from
\li7, we take the lower bound  from (\ref{rfbon}) and the upper bound
from\cite{liglob}, giving
\beq
\label{eq:li7-updown}
\EE{9.0}{-11} < \frac{\li7}{\rm H}_p < \EE{2.8}{-10}.
\eeq

Deuterium is measured in high-redshift QSO absorption line systems via its
isotopic shift from hydrogen. In several absorbers of moderate column
density (Lyman-limit systems), D has been observed in multiple Lyman
transitions\cite{omeara,tosl}. Restricting our attention to the three
most reliable regions\cite{omeara}, we find a weighted mean of
\beq
\label{eq:D_p}
\frac{\rm D}{\rm H}_p = (2.9 \pm 0.3) \times 10^{-5}.
\eeq
It should be noted, however, that the $\chi^2$ per degree of freedom is
rather poor ( $\sim 3.4$), and that the unweighted dispersion of these
data  is $\sim 0.6\times 10^{-5}$. This
already points to the dominance of systematic effects. Observation of D in
systems with higher column density (damped systems) find lower
D/H~\cite{pb}, at a level inconsistent with (\ref{eq:D_p}), further
suggesting that systematic effects dominate the error budget\cite{foscv}.
If  all five available observations are used, we would find D/H = $(2.6
\pm 0.3) \times 10^{-5}$ with an even worse $\chi^2$ per degree of freedom
($\sim 4.3$) and an unweighted dispersion of 0.8.

Because there are no known astrophysical sites for the production of
deuterium, all observed D is assumed to be primordial\cite{rafs}. As a
result, any firm determination of a deuterium abundance establishes an
upper bound on $\eta$ which is robust.  Thus, the recent measurements of
D/H~\cite{omeara} at least provide a lower bound on D/H, D/H $> 2.1
\times 10^{-5} (2 \sigma)$ and hence provide an upper bound to $\eta$,
$\eta_{10} < 7.3$ and $\Omega_B h^2 < 0.027$.

Helium-3 can be measured through its hyperfine emission in the radio band,
and has been observed in H{\small II} regions in our
Galaxy.  These observations find\cite{bbr} that there are no obvious trends
in \he3 with metallicity and location in the Galaxy. There is, however,
considerable scatter in the data by a factor $\sim 2$, some of which may
be real. Unfortunately, the stellar and Galactic evolution of \he3 is not
yet sufficiently well understood to confirm whether \he3 is increasing or
decreasing from its primordial value\cite{vofc}.  Consequently, it is
unclear whether the observed \he3 abundance represents an
upper or lower limit to the primordial value. Therefore, we can not use
\he3 abundance as a constraint. 

By combining the predictions of BBN calculations with the abundances of 
D, \he4, and \li7 discussed above
one can determine the
the 95\% CL region 
$4.9<
\eta_{10} < 6.4$, with the peak value occurring at
$\eta_{10} = 5.6$. This range corresponds to values of
$\Omega_B$ between
\beq
0.018 < \Omega_B h^2 < 0.023
\label{omega3}
\eeq 
with a central value of $\Omega_B h^2 = 0.020$.

If we were to use only the deuterium abundance from Eq. \ref{eq:D_p}, one
obtains the 95\% CL  range
$5.3 <
\eta_{10} < 7.3$, with the peak value occurring at
$\eta_{10} = 5.9$. This range corresponds to values of
$\Omega_B$ between
\beq
0.019 < \Omega_B h^2 < 0.027 
\label{omega2}
\eeq 
with a central value of $\Omega_B h^2 = 0.021$.
As one can see from a comparison with Table 1, these values are
in excellent agreement with determinations from the CMB.

\subsection{Candidates}

\subsubsection{Baryons}

Accepting the dark matter hypothesis, the first choice for a
 candidate should be something we know to exist, baryons.  
Though baryonic dark matter can not be the whole story if $\Omega_m >
0.1$,  the identity of the
 dark matter in galactic halos, which appear to contribute at the 
level of $\Omega \sim 0.05$,  remains an important question needing to be
resolved.  A baryon density of this magnitude is not excluded by 
nucleosynthesis. 
 Indeed we know some of the baryons are dark since $\Omega \la 0.01$ 
in the disk of the galaxy.

It is interesting to note that until recently, there seemed to be some
difficulty in reconciling the baryon budget of the Universe.
By counting the visible contribution to $\Omega$ in stellar populations
and  the X-ray producing hot gas, Persic and Salucci\cite{ps} found only
$\Omega_{\rm vis} \simeq 0.003$. A subsequent accounting by Fukugita,
Hogan and Peebles\cite{fhp} found slightly more ($\Omega \sim 0.02$) by
including the contribution from plasmas in groups and clusters.
At high redshift on the other hand, all of the baryons can be accounted
for. The observed opacity of the Ly $\alpha$ forest in QSO absorption
spectra requires a large baryon density consistent with the
determinations by the CMB and BBN\cite{hae}.

In galactic halos, however, it is quite difficult to hide large
amounts  of baryonic matter\cite{hio12}. Sites for halo baryons that 
have been discussed
include Hydrogen (frozen, cold or hot gas), low mass stars/Jupiters,
remnants of massive stars such as white dwarfs, neutron stars or black
holes.  In almost every case, a serious theoretical or observational
problem is encountered.  

\vskip .1in
\noindent {\it 2.2.1.1 Hydrogen}
\vskip .1in

	A halo predominately made of hydrogen 
(with a primordial admixture of $^4$He) is perhaps the 
simplest possibility.  Hydrogen may however be present 
in a condensed snow-ball like state or in the form of gas.
 Aside from the obvious question of how do these snowballs
 get made, it is possible to show that their existence today 
requires them to be so large as to be gravitationally bound\cite{hio12}.

	Assuming that these objects are electrostatically bound, the 
average density of solid hydrogen is $\rho_s = 0.07$ g cm$^{-3}$ and the
 binding energy per molecule is about 1 eV. Given that the average 
velocity of a snowball is $v \sim 250$ kms$^{-1}$, corresponding to a kinetic
 energy $E_k \sim 600$ eV, snowballs must be collisionless in order to survive.
 Requiring that the collision rate $\Gamma_c  = n_s \sigma v$
 be less than ${t_u}^{-1}$
($t_u$ is the age of the halo $\simeq$ age of the Universe) with 
$n_s = \rho_H/m_s$ and $\sigma = \pi {r_s}^2$ one finds $r_s \ga 2$ cm and
 $m_s \ga 1$ g, assuming a halo density $\rho_H = 1.7 \times 10^{-26}$ g
cm$^{-3}$ (corresponding to $10^{12} M_\odot$ in a radius of 100 kpc). However,
collisionless snowballs also require
 that their formation occur when the overall density $\rho = \rho_H$.
 In this case, snowballs could not have formed later than a
 redshift $(1+z) = 3.5$ or when the microwave background
 temperature was 9.5K.  At this temperature, there is no
 equilibrium between the gaseous and condensed state and 
the snowballs would sublimate. For a snowball to survive, $r_s > 10^{16}$
cm is required making this no longer
 an electrostatically bound object. 

	If snowballs sublimate, then we can consider
 the possibility of a halo composed of cold hydrogen gas. 
 Because the collapse time-scale for the halo ($< 10^9$ yrs) 
is much less than the age of the galaxy, the gas must be in 
hydrostatic equilibrium.  Combining the equation of state
\beq
	P(r) = (2\rho(r)/m_p)kT	
\eeq
where $m_p$ is the proton mass, with
\beq
	dP(r)/dr = -{GM(r)\rho(r) \over r^2}	
\eeq
one can solve for the equilibrium temperature
\beq
	T = {G m_p M(r) \over 4 k r} \simeq 1.3 \times 10^6 K	
\eeq
This is hot gas. As discussed earlier, hot gas is observed through X-ray
emission.  It is easy to show that an entire halo of hot gas would
conflict severely with observations. Cooling of course may occur, but at
the expense of star formation.

\vskip .1in
\noindent {\it 2.2.1.2 Jupiter-like objects}
\vskip .1in
	A very popular candidate for baryonic dark matter 
is a very low mass star or JLO.  These are objects with
 a mass $m < m_o  = 0.08 M_\odot$, the mass necessary to commence
 nuclear burning.  Presumably there is a minimum mass\cite{min} 
$m > m_{min} = (0.004 - 0.007)M_\odot$  based on fragmentation, but the
exact value is very uncertain. The contribution of these objects to the
dark matter in the halo depends on how much mass can one put between
$m_{min}$ and $m_o$ and depends on the initial mass function (IMF) in the
halo.

\begin{figure}[t]
\hspace{0.5truecm}
\epsfxsize=8cm   %width of figure - will enlarge/reduce the figures
\epsfbox{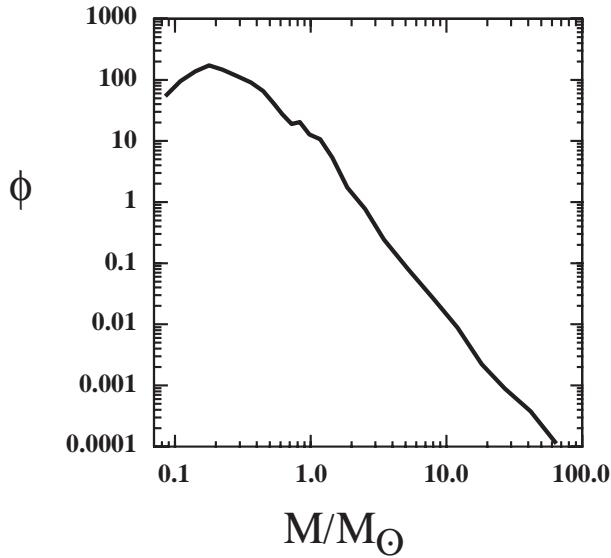}
\caption{{The IMF in the solar neighborhood\protect\cite{scalo}.}}
\label{imf}
\end{figure}

	An IMF is the number of stars formed per unit volume per unit
mass and can be parameterized as
\beq
\phi = Am^{-(1+x)}	
\eeq
 In this parametrization, the Salpeter mass function
 corresponds to x = 1.35.  Because stars are not 
observed with $m < m_o$, some assumptions about the IMF 
for low masses must be made. An example of the observed\cite{scalo} IMF in
the solar neighborhood is shown in Fig. \ref{imf}.

It is possible to use infrared observations\cite{i} to place a lower
limit on the slope, $x$, of the IMF in the halo of galaxies by comparing 
a mass-to-light ratio defined by 
\beq
	Q = \left(\rho_m/\rho_L\right) L_\odot/M_\odot	
\eeq
where the total mass density in JLO's and low mass stars is given by
\beq
	\rho_m = \int_{m_{min}}^{m_G} m \phi dm 	
\eeq
where $m_G = 0.75 M_\odot$ is the mass of a giant.  
The luminosity density given by such a stellar distribution is
\beq
	\rho_L  =    \int_{m_o}^{m_G} L \phi dm + \rho_G	
\eeq
where $L(m)$ is the luminosity of a star of mass $m$. $\rho_G$
  is the contribution to the luminosity density due to giant stars.
The observed\cite{i} lower limits on $Q$ translate to a limit\cite{hio12}
on $x$
\beq
	x > 1.7	
\label{x}
\eeq
with a weak dependence of $m_{min}$.   
 
	Unfortunately, one can not use Eq. (\ref{x}) to exclude JLO's
 since we do not observe an IMF in the halo and it may be different
 from that in the disk.  One can however make a comparison with 
existing observations, none of which show such a steep slope at low
masses. Indeed,  most observations leading to a determination of the IMF
(such as the one shown in Fig. \ref{imf}) show a turn over (or negative
slope). 
	To fully answer the questions regarding JLO's in the halo,
 one needs a better understanding of star formation and the IMF.
For now, postulating the existence of a large fraction of JLO's in halo
is rather ad-hoc.

Despite the theoretical arguments against them, JLO's or massive compact
halo objects (MACHOs) are candidates which are testable by the
gravitational microlensing of stars in a neighboring galaxy such as the
LMC\cite{pac}. By observing millions of stars and examining their
intensity as a function of time, it is possible to determine the
presence of dark objects in our halo. It is expected that during a
lensing event, a star in the LMC will have its  intensity rise in an
achromatic fashion over  a period
$\delta t
\sim 3$ $\sqrt{M/.001 M_\odot}$ days.
Indeed, microlensing candidates have been found\cite{macho}. For low mass
objects, those with $M < 0.1M_\odot$, it appears however that 
the halo fraction
of MACHOs is very small. The relative amount of machos in the halo is
typically expressed in terms of an optical depth.  A halo consisting
100\% of machos  would have an optical depth of $\tau \sim 5 \times
10^{-7}$. The most recent results of the MACHO collaboration\cite{m2} 
indicate that $\tau = 12^{+4}_{-3} \times 10^{-8}$,
corresponding to a  macho halo fraction of about 20\% with a 95\% CL
range of 8 -- 50\% based on 13 -- 17 events. They also exclude a 100\%
macho halo at the 95\% CL. The typical macho mass falls in the range 0.15
-- 0.9 $M_\odot$. The EROS collaboration has set even stronger limits
having observed 5 events toward the LMC and 4 toward the SMC\cite{er}.  
The observed optical depth from EROS1 is 
$\tau = 4^{+10}_{-4} \times 10^{-8}$ and from EROS2 $\tau = 6^{+5}_{-3} \times
10^{-8}$.  They have excluded low mass objects ($M < 0.1M_\odot$) to make up less than
10\% of the halo and objects with $2 \times 10^{-7} M_\odot < M < 1
M_\odot$ to be less than 25\% of the halo at the 95\% CL.

\vskip .1in
\noindent {\it 2.2.1.3 Remnants of Massive Stars}
\vskip .1in
Next one should possibility that the halo is made up
 of the dead stellar remnants of stars whose initial
 masses were $M > 1M_\odot$. Briefly, the problem which arises 
in this context is that since at least 40\% of the stars
 initial mass is ejected, and most of this mass is in the form 
of heavy elements, a large population of these objects would contaminate
 the disk and prevent the existence of extremely low metallicity objects 
($Z \sim 10^{-5}$) which have been observed.  Thus either dust (from
ejecta) or dead remnants would be expected to produce too large a
metallicity\cite{hio12,carr}. Clearly star formation is a very inefficient
mechanism for producing dark matter.  Many generations of stars would
would be required to cycle through their lifetimes to continually
trap more matter in remnants.

	This question has been studied in more detail\cite{ros}.  
By allowing a variable star formation rate, and allowing a great 
deal of flexibility in the IMF, a search for a consistent 
 set of parameters so that the halo could be described primarily
 in terms of dead remnants (in this case white dwarfs) was performed. 
While a consistent set of chemical evolution parameters can be found,
there is no sensible theory to support this choice.
 In such a model however, 
the dark matter is in the form of white dwarfs and the remaining gas
 is is heavily contaminated by Carbon and Nitrogen\cite{ffg}.  Though it
is not excluded, it is hard to understand
 this corner of parameter space as being realistic.

\vskip .1in
\noindent {\it 2.2.1.4 Black Holes}
\vskip .1in
	There are several possibilities for black holes as the dark matter in halos.
  If the black holes are primordial\cite{jane} which have presumably 
formed before nucleosynthesis, they should not be counted
 as baryonic dark matter and therefore do not enter into the
 present discussion.  If the black holes are formed as the final
 stage of star's history and its formation was preceded by mass
 loss or a supernovae, then the previous discussion on remnants
 of massive stars applies here as well.  However, it is also possible
 that the black halos were formed directly from very massive
 stars ($m > 100 M_\odot$?) through gravitational 
instability with no mass loss\cite{carr2}. Though there are limits
due to overheating the disk\cite{lo} and stellar systems\cite{rl}.
 In this case I know of no argument preventing
 a sufficiently large population of massive black
 holes as baryonic dark matter (Of course, now the
 IMF must have $m_{min} \ga 100 M_\odot.$)

\subsubsection{Neutrinos}

	Light neutrinos ($m \le 30 eV$) are 
a long-time standard when it comes to
 non-baryonic dark matter\cite{ss}.  Light neutrinos produce 
structure on large scales, and the natural (minimal) scale for
 structure clustering is given in Eq. (\ref{mj}).  Hence neutrinos
 offer the natural possibility for large scale structures\cite{nu1,nu2} 
including filaments and voids.  Light neutrinos
 are, however, ruled out as a dominant form of dark matter because they 
 produce too much large scale structure\cite{nu3}.
  Because the smallest non-linear structures have mass scale $M_J$ and 
the typical galactic mass scale is $\simeq 10^{12} M_\odot$, galaxies must 
fragment out of the larger pancake-like objects.  The problem with
such a scenario is that galaxies form late\cite{nu2,nu4} 
 ($z \le 1$) whereas
 quasars and galaxies are seen out to redshifts $z \ga 6$. 

In the standard model, the absence of a right-handed neutrino state precludes
the existence of a neutrino mass (unless one includes non-renormalizable
lepton number violating interactions such HHLL).  By adding a
right-handed  state $\nu_R$, it is possible to generate a Dirac mass
for the neutrino, $m_\nu = h_\nu v/\sqrt{2}$,
as is the case for the charged lepton masses, where $h_\nu$ is the
neutrino Yukawa coupling constant, and $v$ is the Higgs expectation
value.  It is also possible to generate a Majorana mass for the neutrino
when in addition to the Dirac mass term, $m_\nu \bar{\nu_R} \nu_L$, a
term $M
\nu_R \nu_R$ is included. If $M \gg m_\nu$, the see-saw mechanism produces
two mass eigenstates given by $m_{\nu_1} \sim m_\nu^2/M$ which is very
light, and $m_{\nu_2} \sim M$ which is heavy.  The state $\nu_1$ is a
potential hot dark matter candidate as $\nu_2$ is in general not stable.

The simplicity of the standard big bang model allows one to
compute in a straightforward manner the relic density of any
stable particle if that particle was once in thermal equilibrium
with the thermal radiation bath.  At early times, neutrinos were
kept in thermal equilibrium by their weak interactions with electrons 
and positrons. As we saw in the case of the $\beta$-interaction used in
BBN, one can estimate the thermally averaged  low-energy weak interaction
scattering cross section
\beq
 \langle \sigma v \rangle\; { \sim }~g^4 T^{ 2} /m_{W}^4
\eeq
for $T \ll m_W$. Recalling that the number density scales as 
$n \propto T^3$, we can compare the weak interaction rate 
$\Gamma \sim n \langle \sigma v \rangle$, with the expansion rate 
given by eqs. (\ref{fried}) with 
\beq
 \rho {} = \left( \sum_B g_{B} + {7 \over 8} \sum_F  g_{F} \right)
   {\pi^{ 2} \over 30}  T^{4} \equiv    {\pi^{ 2} \over 30}\, N(T)
\, T^{4} 
\label{rho1}
\eeq
Neutrinos will be in 
equilibrium when $\Gamma_{\rm wk} > H$ or
\beq
  T^3 > \sqrt{8 \pi^3 N /90}\,\,\, m_{W}^4/M_{P}
\eeq
where $M_{P} = G_{N}^{-1/2} =  1.22 \times 10^{19}$ GeV is the Planck
mass.  For $N = 43/4$ (accounting for  photons, electrons, positrons 
and three neutrino flavors) we see that equilibrium is maintained at
temperatures greater than ${\mathcal O}(1)$ MeV (for a more accurate 
calculation see~\cite{Enqvist:gx}).

The decoupling scale of ${\mathcal O}(1)$ MeV has an important
consequence  on the final relic density of massive neutrinos. Neutrinos
more massive  than 1 MeV will begin to annihilate prior to decoupling,
and while in  equilibrium, their number density will become exponentially
suppressed.  Lighter neutrinos decouple as radiation on the other hand,
and hence do  not experience the suppression due to annihilation.
Therefore, the  calculations of the number density of light ($m_\nu \la
1$ MeV) and  heavy ($m_\nu \ga 1$ MeV) neutrinos differ substantially.

\begin{figure}[t]
\hspace{0.5truecm}
\epsfxsize=8cm   %width of figure - will enlarge/reduce the figures
\epsfbox{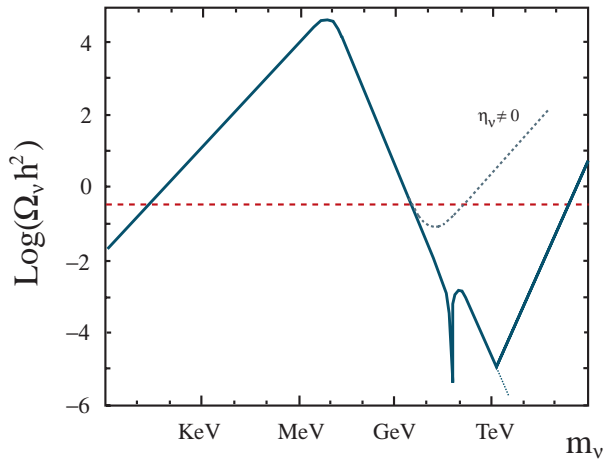}
\caption{{Summary plot\protect\cite{kko} of the relic density of Dirac
neutrinos  (solid) including a possible neutrino asymmetry of $\eta_\nu =
5\times 10^{-11}$ (dotted).}}
\label{nu}
\end{figure}

The number of density of light neutrinos with $m_\nu \la 1$ MeV can be 
expressed at late times as
\beq
  \rho_\nu  = m_\nu Y_\nu n_\gamma  
\label{rhonus}
\eeq
where $Y_\nu = n_\nu/n_\gamma$ is the density of $\nu$'s relative to 
the density of photons, which today is 411 photons per cm$^3$. It is 
easy to show that in an adiabatically expanding universe $Y_\nu = 
3/11$. This suppression is a result of the $e^+ e^-$ annihilation 
which occurs after neutrino decoupling and heats the photon bath 
relative to the neutrinos.
In order to obtain an age of the Universe, $t > 12$ Gyr, one 
requires that the matter component is constrained by
\beq
  \Omega h^2 \le 0.3.
\label{omegabound}
\eeq  
From this one finds the strong constraint (upper bound) on Majorana
neutrino masses\cite{cows}:
\beq
m_{\rm tot} =   \sum_\nu  m_\nu   \la 28 {\rm eV}.
\label{ml1}
\eeq
where the sum runs over neutrino mass eigenstates. The limit for Dirac
neutrinos depends on the interactions of the right-handed states (see
discussion below).  Given the discussion of the CMB results in the
previous section, one could make a case that the limit on $\Omega h^2$
should be reduced by a factor of 2, which would translate in to a limit
of 14 eV on the sum of the light neutrino masses. As one can see, even
very small  neutrino masses of order 1 eV, may contribute substantially
to the  overall relic density. The limit (\ref{ml1}) and the
corresponding  initial rise in
$\Omega_\nu h^2$ as a function of $m_\nu$ is  displayed in the
Figure~\ref{nu} (the low mass end with 
$m_\nu \la 1$ MeV).

Combining the rapidly improving data on key cosmological 
parameters with the better statistics from large redshift surveys 
has made it possible to go a step forward along this path. It is 
now possible to set stringent limits on the light neutrino mass
density $\Omega_\nu h^2$, and hence on the neutrino mass based on
the power spectrum of the Ly 
$\alpha$ forest\cite{strong}, $m_{\rm tot} < 5.5$ eV, and the limit 
is even stronger if the total matter density, $\Omega_m$ is less 
than 0.5. Adding additional observation constraints from the CMB and
galaxy clusters drops this limit\cite{wang} to 2.4 eV. This limit has
recently been improved by the 2dF Galaxy redshift\cite{2dF} 
survey by comparing the derived power spectrum of fluctuations
with structure formation models. 
 Focussing on the the 
presently favoured $\Lambda$CDM model, the neutrino mass bound becomes
$m_{\rm tot} < 1.8 $ eV for $\Omega_m < 0.5$.  When even more constraints
such as HST Key project data,  supernovae type Ia data, and BBN are
included\cite{Lewis} the limit can be pushed to $m_{\rm tot} < 0.3 $ eV.

The calculation of the relic density for neutrinos more massive than
$\sim 1$ MeV, is substantially more involved. The relic density is now
determined by the freeze-out of neutrino annihilations which occur at
$T \la m_\nu$, after annihilations have begun to seriously reduce their
number density\cite{lw}. The annihilation rate is given by
\beq
\Gamma_{ann} = \langle \sigma v \rangle_{ann} n_\nu \sim
\frac{m_\nu^2}{m_Z^4} (m_\nu T)^{3/2} e^{-m_\nu/T}
\label{annrate}
\eeq
where we have assumed, for example, that the annihilation cross section
is dominated by $\nu {\bar \nu} \rightarrow f {\bar f}$ via $Z$-boson
exchange\footnote{While this is approximately true for Dirac neutrinos,
the annihilation cross section of Majorana neutrinos is $p$-wave
suppressed and is proportional of the final state fermion masses 
rather than $m_\nu$.} 
and $\langle \sigma v \rangle_{ann} \sim m_\nu^2/m_Z^4$. When the 
annihilation rate becomes slower than the expansion rate of the Universe
the annihilations freeze out and the relative abundance of neutrinos
becomes fixed. 
 For particles
which annihilate through approximate weak scale interactions, this occurs
when $T \sim m_\chi /20$.  The number density of neutrinos is tracked
by a Boltzmann-like equation,
\beq
{dn \over dt} = -3{{\dot R} \over R} n - \langle \sigma v \rangle (n^2 -
n_0^2)
\label{rate}
\eeq
where $n_0$ is the equilibrium number density of neutralinos.
By defining the quantity $f = n/T^3$, we can rewrite this equation in terms of $x$,
as
\beq
{df \over dx} = m_\nu \left( {8 \pi^3 \over 90}G_N N \right)^{1/2}
(f^2 - f_0^2)
\label{rate2}
\eeq
The solution to this equation at late times (small $x$) yields a constant value of
$f$, so that $n \propto T^3$.

\begin{figure}
\begin{center}
\hspace{0.5truecm}
\epsfxsize=8cm   %width of figure - will enlarge/reduce the figures
\epsfbox{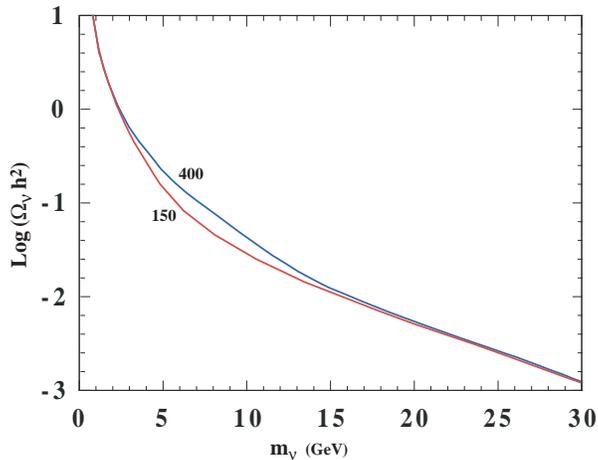}
\caption{The relic density of heavy Dirac neutrinos due to 
annihilations\protect\cite{wso}. The curves are labeled by
the assumed quark-hadron phase transition temperature in MeV.}
\label{swo}       
\end{center}
\end{figure}

Roughly, the solution to the Boltzmann equation goes as $Y_\nu \sim f \sim (m
\langle
\sigma v
\rangle_{ann} )^{-1}$ and hence $\Omega_\nu h^2 \sim {\langle \sigma v
\rangle_{ann}}^{-1}$, so that parametrically $\Omega_\nu h^2  \sim
1/{m_\nu^2}$. As a result, the constraint (\ref{omegabound}) now leads to
a {\em lower}  bound\cite{lw,ko,wso} on the neutrino mass, of about
$m_\nu \ga 3-7$  GeV, depending on whether it is a Dirac or Majorana
neutrino.  This bound and the corresponding downward trend $\Omega_\nu
h^2 \sim 1/m^2_\nu$ can again be seen in Figure~\ref{nu}. The result of a
more detailed calculation is shown in
Figure~\ref{swo}~\cite{wso} for the case of a Dirac neutrino. 
The two curves show the slight sensitivity on the temperature scale
associated with the quark-hadron transition. The result for a Majorana
mass neutrino is qualitatively similar. 
Indeed, any particle with roughly weak scale cross-sections will tend to
give an interesting  value of  $\Omega h^2 \sim 1$.

The deep drop in $\Omega_\nu h^2$, visible in Figure~\ref{nu}
at around $m_\nu = M_Z/2$, is due to a very strong annihilation 
cross section at $Z$-boson pole. For yet higher neutrino masses the 
$Z$-annihilation channel cross section drops as $\sim 1/m_\nu^2$, 
leading to a brief period of an increasing trend in $\Omega_\nu h^2$. 
However, for $m_\nu \ga m_W$ the cross section regains its parametric 
form $\langle \sigma v \rangle_{ann} \sim m_\nu^2$ due to the opening up
of a new annihilation channel to $W$-boson pairs\cite{Enqvist:1988we}, 
and the density drops again as $\Omega_\nu h^2 \sim 1/m^2_\nu$. 
The tree level $W$-channel cross section breaks the unitarity at 
around ${\mathcal O}({\rm few})$ TeV~\cite{Enqvist:yz} however, and the
full  cross section must be bound by the unitarity
limit\cite{Griest:1989wd}. This behaves again as $1/m_\nu^2$, whereby
$\Omega_\nu h^2$  has to start increasing again, until it becomes too
large again at 200-400 TeV~\cite{Griest:1989wd,Enqvist:yz} (or perhaps
somewhat  earlier as the weak interactions become strong at the unitarity
breaking scale).

If neutrinos are Dirac particles, and 
have a nonzero asymmetry  the relic density could 
be governed by the asymmetry rather than by the annihilation cross section. 
Indeed, it is easy to see that the neutrino mass density corresponding 
to the asymmetry $\eta_\nu \equiv (n_\nu - n_{\bar \nu})/n_\gamma$ is 
given by\cite{ho}
\beq
  \rho = m_\nu \eta_\nu n_\gamma ,
\eeq
which implies
\beq
  \Omega_\nu h^2 \simeq 0.004 \,\eta_{\nu 10}\, (m_\nu/{\rm GeV}).
\eeq
where $\eta_{\nu 10}\equiv 10^{10}\eta_\nu$.
The behaviour of the energy density of neutrinos
with an asymmetry is shown by the dotted line in the Figure~\ref{nu}.
At  low $m_\nu$, the mass density is dominated by the symmetric, 
relic abundance of both neutrinos and antineutrinos which have already
frozen out. At higher values  of $m_\nu$, the annihilations suppress the
symmetric part of the relic  density until $\Omega_\nu h^2$
eventually becomes  dominated by the linearly increasing asymmetric
contribution. In the  figure, we have assumed an asymmetry of
$\eta_\nu \sim 5 \times  10^{-11}$ for neutrinos with standard weak 
interaction strength. In this case, $\Omega_\nu h^2$ begins to rise 
when $m_\nu \ga 20$ GeV.  Obviously, the bound (\ref{omegabound}) is 
saturated for $m_\nu = 75 \, {\rm GeV}/\eta_{\nu 10}$.

Based on the leptonic and invisible width of the $Z$ boson, 
experiments at LEP have determined that the number of neutrinos is 
$N_\nu = 2.9841 \pm 0.0083$~\cite{RPP}. Conversely, any new physics 
must fit within these brackets, and thus LEP excludes additional 
neutrinos (with standard weak interactions) with masses $m_\nu 
\la 45$ GeV.  Combined with the limits displayed in Figures~\ref{nu} 
and \ref{swo}, we see that the mass density of ordinary heavy
neutrinos is bound to be very small, $\Omega_\nu {h}^2 < 0.001$ for
masses  $m_\nu > 45$ GeV up to $m_\nu \sim {\mathcal O}(100)$ TeV. 
Lab constraints for 
Dirac neutrinos are available\cite{dir}, excluding neutrinos 
with masses between
10 GeV and 4.7 TeV. This is significant, since it precludes the possibility 
of neutrino dark matter based on an asymmetry between $\nu$ and ${\bar \nu}$ 
\cite{ho}. Majorana neutrinos are excluded as {\em dark matter}
since $\Omega_\nu {h_o}^2 < 0.001$ for $m_\nu > 45$ GeV and are thus
cosmologically  uninteresting.

A bound on neutrino masses even stronger than Eqn.~(\ref{ml1}) can be 
obtained from the recent observations of active-active mixing in both 
solar- and atmospheric neutrino experiments. The inferred evidence for 
$\nu_\mu-\nu_\tau$ and $\nu_e-\nu_{\mu,\tau}$ mixings are on the scales 
$m_\nu^2 \sim 1-10 \times 10^{-5}$ and $m_\nu^2  \sim 2-5 \times 10^{-3}$. 
When combined with the upper bound on the electron-like neutrino mass
$m_{\nu} <  2.8$ eV~\cite{Mainz}, and the LEP-limit on the number of
neutrino  species, one finds the constraint on the sum of neutrino masses:
\beq
0.05~{\rm eV} \la m_{\rm tot} \la 8.4~\rm eV.
\eeq
Conversely, the experimental and observational data then implies that 
the cosmological energy density of all light, weakly interacting 
neutrinos can be restricted to the range 
\beq
0.0005 \la \Omega_\nu h^2 \la 0.09.
\label{range}
\eeq
Interestingly there is now also a lower bound due to the fact that at 
least one of the neutrino masses has to be larger than the scale $m^2 
\sim 10^{-3}$ eV$^2$ set by the atmospheric neutrino data.
Combined with the results on relic mass density of neutrinos and the
LEP limits, the bound~(\ref{range}) implies that the ordinary weakly 
interacting neutrinos, once the standard dark matter 
candidate\cite{ss}, can be ruled out completely as a dominant 
component of the dark matter.

If instead, we consider right-handed neutrinos, we have new possibilities.
Right-handed interactions are necessarily weaker than standard
left-handed  interactions implying that right-handed neutrinos
decouple early and today are at a reduced  temperature relative to
$\nu_L$~\cite{oss} 
\beq
({T_\chi \over T_\gamma})^3 = {43 \over 4 N(T_d)}
\label{decx}
\eeq
As such, for ${T_d}_R \gg 1$ MeV, $n_{\nu_R}/n_{\nu_L} = 
(T_{\nu_R}/T_{\nu_L})^3 \ll 1$. Thus the abundance of right-handed
neutrinos can be written as 
\beq
Y_{\nu_R} = {n_{\nu_R} \over n_\gamma} = ({3 \over 11})
({T_{\nu_R} \over T_{\nu_L}})^3 \ll {3 \over 11}
\eeq
In this case, the previous bound (\ref{ml1}) on neutrino masses is
weakened.
 For a suitably large scale for the right-handed interactions,
 right-handed neutrino masses may be as large as a few keV \cite{ot}.
Such neutrinos make excellent warm dark matter candidates, albeit the 
viable mass range for galaxy formation is quite restricted\cite{warm}.

\subsubsection{Axions}

Due to space limitations, the discussion of this candidate will be very
brief.
Axions are
pseudo-Goldstone bosons which arise in solving the strong CP
problem\cite{ax1,ax2} via a global U(1) Peccei-Quinn symmetry.  The
invisible axion\cite{ax2} is associated with the flat direction of the
spontaneously broken PQ symmetry.  Because the PQ symmetry is also
explicitly broken (the CP violating $\theta F {\widetilde F}$ coupling is
not PQ invariant) the axion picks up a small mass similar to pion picking
up a mass when chiral symmetry is broken.  We can expect that $m_a  \sim
m_\pi f_\pi /f_a$  where $f_a$, the axion decay constant, is the vacuum
expectation value of the PQ current and can be taken to be quite large. 
If we write the axion field as $a = f_a \theta$, near the minimum, the
potential produced by QCD instanton effects looks like $V \sim m_a^2
\theta^2 f_a^2$.  The axion equations of motion lead to
a relatively stable oscillating solution.  The energy density stored in
the oscillations exceeds the critical density\cite{axden}  unless $f_a 
\la 10^{12}$ GeV.

	Axions may also be emitted stars and supernova\cite{raff}.
In supernovae, axions are produced via nucleon-nucleon bremsstrahlung with
a coupling $g_AN \propto m_N/f_a$.  As was noted above the cosmological
density limit requires $f_a \la 10^{12}$  GeV.  Axion emission from red
giants imply\cite{dss}   $ f_a  \ga 10^{10}$ GeV (though this limit
depends on an adjustable axion-electron coupling), the supernova limit
requires\cite{sn}   $ f_a \ga 2 \times 10^{11}$   GeV for naive quark
model couplings of the axion to nucleons.   
Thus only a narrow window exists for the axion as a viable dark matter
candidate.

\section{Lecture 3: Supersymmetric Dark Matter}

Although there are many reasons for
considering supersymmetry as a candidate
extension to the standard model of strong,
weak and electromagnetic interactions\cite{reviews}, one of
the most compelling is its role in
understanding the hierarchy problem\cite{hierarchy} namely, 
why/how is
$m_W \ll M_P$.  One might think naively that it would be
sufficient to set $m_W \ll M_P$ by hand. However, radiative
corrections tend to destroy this hierarchy. For example,
one-loop diagrams generate
\beq
\delta m^2_W = \mathcal {O}\left({\alpha\over\pi}\right)~\Lambda^2 \gg m^2_W
\label{four}
\eeq
where $\Lambda$ is a cut-off representing the appearance of new physics, and
the
inequality in (\ref{four}) applies if $\Lambda\sim 10^3$ TeV, and even 
more so if  $\Lambda \sim m_{GUT} \sim
10^{16}$
GeV or $ \sim M_P \sim 10^{19}$ GeV. If the radiative corrections to a 
physical
quantity
are much larger than its measured values, obtaining the latter requires
strong
cancellations, which in general require fine tuning of the bare input
parameters.
However, the necessary cancellations are natural in supersymmetry, where 
one has
equal
numbers of bosons  and fermions  with equal couplings, so that
(\ref{four})
is replaced by
\beq
\delta m^2_W = \mathcal {O}\left({\alpha\over\pi}\right)~\vert m^2_B - 
m^2_F\vert~.
\label{five}
\eeq
The residual radiative correction is naturally small if
$
\vert m^2_B - m^2_F\vert \la 1~{\rm TeV}^2
$.

In order
to justify the absence of interactions
which can be responsible for extremely rapid proton decay, it
is common in the minimal supersymmetric standard model (MSSM)
to assume the conservation of R-parity.  If R-parity, which
distinguishes between ``normal" matter and the  supersymmetric
partners and can be defined in terms of baryon, lepton and spin
as $R = (-1)^{3B + L + 2S}$, is unbroken, there is at least one 
supersymmetric particle (the lightest supersymmetric particle or LSP)
which must be stable.  Thus, the minimal model contains the fewest number of new
particles and interactions necessary to make a consistent theory.

There are very strong constraints, however, forbidding the existence of stable or
long lived particles which are not color and electrically neutral\cite{EHNOS}. Strong
and electromagnetically interacting LSPs would become bound with normal matter
forming anomalously heavy isotopes. Indeed, there are very strong upper limits on the
abundances, relative to hydrogen, of nuclear isotopes\cite{isotopes},
$n/n_H \la 10^{-15}~~{\rm to}~~10^{-29}
$
for 1 GeV $\la m \la$ 1 TeV. A strongly interacting stable relic is expected
to have an abundance $n/n_H \la 10^{-10}$
with a higher abundance for charged particles.

There are relatively few supersymmetric candidates which are not colored and
are electrically neutral.  The sneutrino\cite{snu} is one possibility,
but in the MSSM, it has been excluded as a dark matter candidate by
direct\cite{dir} and indirect\cite{indir} searches.  In fact, one can set
an accelerator based limit on the sneutrino mass from neutrino counting, 
$m_{\tilde\nu}\ga$ 44.7 GeV \cite{EFOS}. In this case, the direct relic
searches in
underground low-background experiments require  
$m_{\tilde\nu}\ga$ 20 TeV~\cite{dir}. Another possibility is the
gravitino which is probably the most difficult to exclude. 
I will concentrate on the remaining possibility in the MSSM, namely the
neutralinos.

\subsection{Parameters}

The most general version of the MSSM, despite its minimality in particles and
interactions contains well over a hundred new parameters. The study of such a model
would be untenable were it not for some (well motivated) assumptions.
These have to do with the parameters associated with supersymmetry breaking.
It is often assumed that, at some unification scale, all of the gaugino masses
receive a common mass, $m_{1/2}$. The gaugino masses at the weak scale are
determined by running a set of renormalization group equations.
Similarly, one often assumes that all scalars receive a common mass, $m_0$,
at the GUT scale. These too are run down to the weak scale. The remaining 
parameters of importance involve the Higgs sector.  There is the Higgs mixing mass
parameter, $\mu$, and since there are two Higgs doublets in the MSSM, there are 
two vacuum expectation values. One combination of these is related to the $Z$ mass,
and therefore is not a free parameter, while the other combination, the ratio of the
two vevs, $\tan \beta$, is free. 

\begin{figure}
\begin{center}
\hspace{0.1truecm}
\epsfxsize=9cm   %width of figure - will enlarge/reduce the figures
\epsfbox{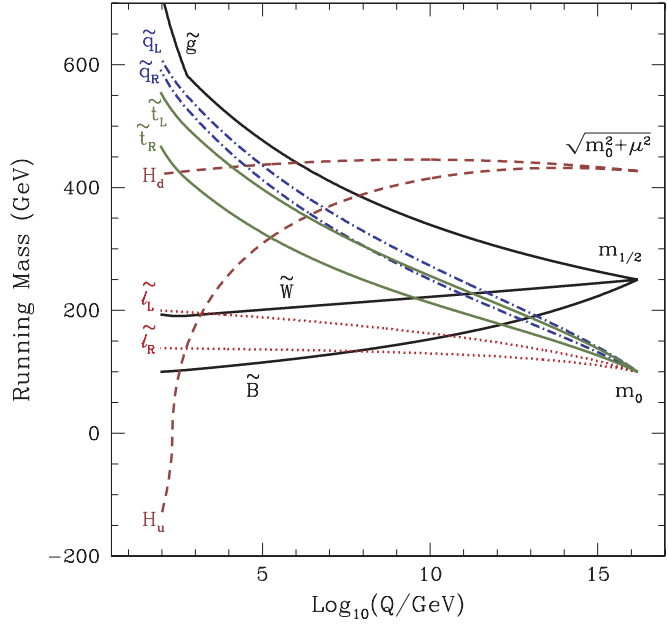}
\caption{RG evolution of the mass parameters in the CMSSM.}
\label{running}       
\end{center}
\end{figure}

If the supersymmetry breaking Higgs soft masses are also unified at the GUT scale
(and take the common value $m_0$), then $\mu$ and the physical Higgs masses at the
weak scale are determined by electroweak vacuum conditions ($\mu$ is
determined up to a sign). This scenario is often referred to as the
constrained MSSM or CMSSM. Once these parameters are set, the entire
spectrum of sparticle masses at the weak scale can be
calculated\footnote{There are in fact, additional parameters: the
supersymmetry-breaking tri-linear masses
$A$ (also assumed to be unified at the GUT scale) as well as two CP
violating phases
$\theta_\mu$ and $\theta_A$.}. 
In Fig. \ref{running}, an example of the running of the mass parameters
in the CMSSM is shown.  Here, we have chosen $m_{1/2} = 250$ GeV, $m_0 = 100$
GeV, $\tan \beta = 3$, $A_0 = 0$, and $\mu < 0$.
Indeed, it is rather amazing that from so few input parameters, all of the
masses of the supersymmetric particles can be determined. 
The characteristic features that one sees in the figure, are for example, that
the colored sparticles are typically the heaviest in the spectrum.  This is
due to the large positive correction to the masses due to $\alpha_3$ in the
RGE's.  Also, one finds that the ${\widetilde B}$ (the partner of the
$U(1)_Y$ gauge boson), is typically the lightest sparticle.  But most
importantly, notice that one of the Higgs mass$^2$, goes negative
triggering electroweak symmetry breaking\cite{rewsb}. 
(The negative sign in the figure refers to the sign of the
mass$^2$, even though it is the mass of the sparticles which are
depicted.)

\subsection{Neutralinos}

 There are four neutralinos, each of which is a  
linear combination of the $R=-1$ neutral fermions\cite{EHNOS}: the wino
$\tilde W^3$, the partner of the
 3rd component of the $SU(2)_L$ gauge boson;
 the bino, $\tilde B$;
 and the two neutral Higgsinos,  $\tilde H_1$ and $\tilde H_2$.
Assuming gaugino mass universality at the  GUT scale, the identity and
mass of the LSP are determined by the gaugino mass $m_{1/2}$, 
$\mu$, and  $\tan \beta$. In general,
neutralinos can  be expressed as a linear combination
\begin{equation}
	\chi = \alpha \tilde B + \beta \tilde W^3 + \gamma \tilde H_1 +
\delta
\tilde H_2
\end{equation}
The solution for the coefficients $\alpha, \beta, \gamma$ and $\delta$
for neutralinos that make up the LSP 
can be found by diagonalizing the mass matrix
\beq
      ({\tilde W}^3, {\tilde B}, {{\tilde H}^0}_1,{{\tilde H}^0}_2 )
  \left( \begin{array}{cccc}
M_2 & 0 & {-g_2 v_1 \over \sqrt{2}} &  {g_2 v_2 \over \sqrt{2}} \\
0 & M_1 & {g_1 v_1 \over \sqrt{2}} & {-g_1 v_2 \over \sqrt{2}} \\
{-g_2 v_1 \over \sqrt{2}} & {g_1 v_1 \over \sqrt{2}} & 0 & -\mu \\
{g_2 v_2 \over \sqrt{2}} & {-g_1 v_2 \over \sqrt{2}} & -\mu & 0 
\end{array} \right) \left( \begin{array}{c} {\tilde W}^3 \\
{\tilde B} \\ {{\tilde H}^0}_1 \\ {{\tilde H}^0}_2 \end{array} \right)
\eeq
where $M_1 (M_2)$ is a soft supersymmetry breaking
 term giving mass to the U(1) (SU(2))  gaugino(s).
  In a unified
 theory $M_1 = M_2 = m_{1/2}$ at the unification scale (at the weak scale,
$ M_1
\simeq {5 \over 3}  {\alpha_1 \over \alpha_2}  M_2	$).   As one can see, 
the coefficients
$\alpha, \beta, \gamma,$ and $\delta$ depend only on
$m_{1/2}$, $\mu$, and $\tan \beta$.

\begin{figure}[th]
	\centering
	\epsfxsize=10cm
\epsfbox{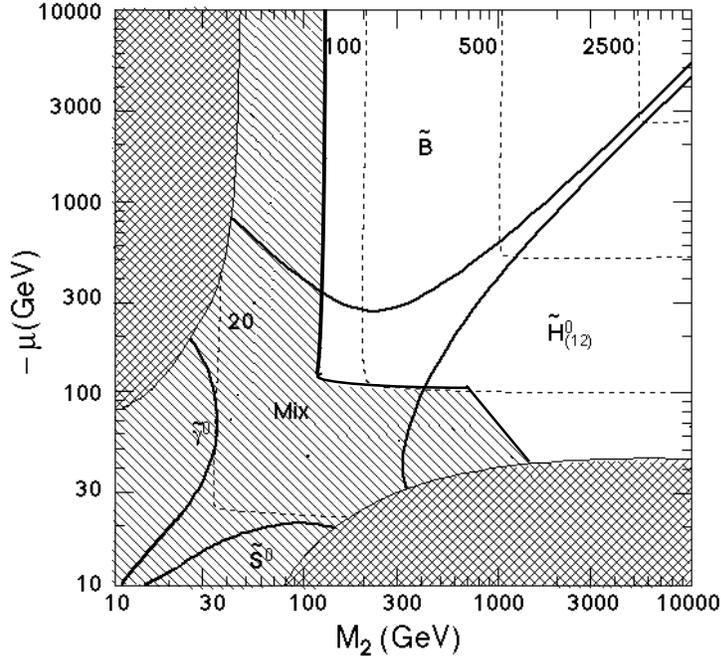}
	\caption{Mass contours and composition of nearly pure LSP states in the MSSM
\protect\cite{osi3}.}
	\label{osi399}
\end{figure}

In Figure \ref{osi399} \cite{osi3}, regions in
the $M_2, \mu$  plane with $\tan\beta = 2$ are shown in which the LSP
is one of several nearly pure states, the photino, $\tilde \gamma$, the
bino,
$\tilde B$, a symmetric combination of the Higgsinos, 
$\tilde{H}_{(12)}$, or the Higgsino, 
$\tilde{S} = \sin \beta {\tilde H}_1 + \cos \beta {\tilde
H}_2$. The dashed lines show the LSP mass contours.
 The cross hatched regions correspond to parameters giving
  a chargino ($\tilde W^{\pm}, \tilde H^{\pm}$) state 
with mass $m_{\tilde \chi} \leq 45 GeV$ and as such are 
excluded by LEP\cite{lep2}.
This constraint has been extended by LEP\cite{LEPsusy} and is shown by
the  light shaded region and corresponds to regions where the chargino
mass is $\la 103.5$ GeV. The newer limit does not extend deep into the
Higgsino region because of the degeneracy between the chargino and
neutralino.
 Notice that the parameter space is dominated by the  
$\tilde B$ or $\tilde H_{12}$
 pure states and that the photino 
 only occupies a small fraction of the parameter space,
 as does the Higgsino combination $\tilde S$. Both of these
light states are experimentally excluded.

\subsection{The Relic Density}

The relic abundance of LSP's is 
determined by solving
the Boltzmann
 equation for the LSP number density in an expanding Universe.
 The technique\cite{wso} used is similar to that for computing
 the relic abundance of massive neutrinos\cite{lw}.
The relic density depends on additional parameters in the MSSM beyond $m_{1/2},
\mu$, and $\tan \beta$. These include the sfermion masses, $m_{\tilde f}$ and the
Higgs pseudo-scalar mass, $m_A$, derived from $m_0$ (and $m_{1/2}$). To
determine the relic density it is necessary to obtain the general
annihilation cross-section for neutralinos.  In much of the parameter
space of interest, the LSP is a bino and the annihilation proceeds mainly
through sfermion exchange. Because of the p-wave suppression associated
with Majorana fermions, the s-wave part of the annihilation cross-section
is suppressed by the outgoing fermion masses.  This means that it is
necessary to expand the cross-section to include p-wave corrections which
can be expressed as a term proportional to the temperature if neutralinos
are in equilibrium. Unless the neutralino mass happens to lie near near a
pole, such as $m_\chi \simeq$
$m_Z/2$ or $m_h/2$, in which case there are large contributions to the
annihilation through direct $s$-channel resonance exchange, the dominant
contribution to
the $\tilde{B} \tilde{B}$ annihilation cross section comes from crossed
$t$-channel sfermion exchange.

Annihilations in the early
Universe continue until the annihilation rate
$\Gamma
\simeq \sigma v n_\chi$ drops below the expansion rate. The calculation
of the neutralino relic density proceeds in much the same way as
discussed above for neutrinos with the appropriate substitution of  the
cross section.
The final neutralino relic density expressed as a fraction of the critical
energy density  can be written as\cite{EHNOS}
\beq
\Omega_\chi h^2 \simeq 1.9 \times 10^{-11} \left({T_\chi \over
T_\gamma}\right)^3 N_f^{1/2} \left({{\rm GeV} \over ax_f + {1\over 2} b
x_f^2}\right)
\label{relic}
\eeq 
where $(T_\chi/T_\gamma)^3$ accounts for the subsequent reheating of the
photon temperature with respect to $\chi$, due to the annihilations of
particles with mass $m < x_f m_\chi$ \cite{oss}. The subscript $f$ refers to
values at freeze-out, i.e., when annihilations cease. The coefficients $a$ and $b$
are related to the partial wave expansion of the cross-section, $\sigma v = a + b x +
\dots $. Eq. (\ref{relic} ) results in a very good approximation to the relic density
expect near s-channel annihilation poles,  thresholds and in regions where the LSP is
nearly degenerate with the next lightest supersymmetric particle\cite{gs}.

When there are several
particle species $i$, which are nearly degenerate in mass,
co-annihilations are important. In this case\cite{gs}, the rate equation
(\ref{rate}) still applies, provided $n$ is
interpreted as the total number density, 
\beq
n \equiv \sum_i n_i \;,
\label{n}
\eeq
$n_0$ as the total equilibrium number density, 
\beq
n_0 \equiv  \sum_i n_{0,i} \;,
\label{neq}
\eeq
and the effective annihilation cross section as
\beq
\langle\sigma_{\rm eff} v_{\rm rel}\rangle \equiv
\sum_{ij}{ n_{0,i} n_{0,j} \over n_0^2}
\langle\sigma_{ij} v_{\rm rel}\rangle \;.
\label{sv2}
\eeq
In eq.~(\ref{rate2}),  $m_\chi$ is now understood to be the mass of the
lightest sparticle under consideration.

 Note that this implies that the
ratio of relic densities computed with and without coannihilations is, 
roughly,
\begin{equation}
  \label{eq:R}
 R\equiv{\Omega^0\over\Omega} 
 \approx \left({\hat\sigma_{\rm eff}\over\hat\sigma_0}\right)
\left({x_{ f}\over x_{ f}^{0}}\right),
\end{equation}
where $\hat\sigma\equiv a + b x/2$ and sub- and superscripts 0 denote
quantities computed ignoring coannihilations.  The ratio ${x_{
    f}^0 / x_{f}}\approx 1+x_{f}^0 \ln (g_{\rm
  eff}\sigma_{\rm eff}/g_1\sigma_0)$, where 
$g_{\rm eff}\equiv\sum_i g_i (m_i/m_1)^{3/2}e^{-(m_i-m_1)/T}$.
For the case of three degenerate slepton NLSPs \cite{efo}, $g_{\rm eff}=\sum_i g_i =8$ and 
${x_{f}^0 / x_{ f}}\approx 1.2$.  The effects of co-annihilations are
discussed below.

\subsection{Phenomenological and Cosmological Constraints}

For the cosmological limits on the relic density I will assume
\begin{equation}   
0.1 \; \le \; \Omega_\chi h^2 \; \le \; 0.3. 
\label{Omegachi}
\end{equation}
The upper limit being a conservative bound based 
only on the lower limit to the
age of the Universe of 12 Gyr. Indeed, most analyses indicate that
$\Omega_{\rm matter} \la 0.4 - 0.5$ and thus it is very likely that
$\Omega_\chi h^2 < 0.2$ (cf. the CMB results in Table 1). One should note
that smaller values of
$\Omega_\chi h^2$  are allowed,
since it is quite possible that some of the cold dark matter might not
consist of LSPs.

The calculated relic density is found to have a relevant cosmological 
density over a wide range of susy parameters. For all values of $\tan \beta$, there
is a `bulk' region with relatively low values of $m_{1/2}$ and $m_0$ where
$0.1 < \Omega_\chi h^2 < 0.3$. However there are a number of regions at large values
of $m_{1/2}$ and/or $m_0$ where the relic density is still compatible with the
cosmological constraints. At large values of $m_{1/2}$, the lighter stau, becomes
nearly degenerate with the neutralino and co-annihilations between these particles
must be taken into account\cite{efo,moreco}. 
For non-zero values of $A_0$, there are new regions 
for which $\chi-{\tilde t}$  coannihilations are important\cite{stopco}. At large
$\tan \beta$, as one increases
$m_{1/2}$, the pseudo-scalar mass, $m_A$ begins to drop so that there is 
a wide funnel-like region (at all values of $m_0$) such that $2m_\chi \approx m_A$ and
s-channel annihilations become important\cite{funnel,EFGOSi}. Finally, there is a
region at very high $m_0$ where the value of $\mu$ begins to fall and the LSP
becomes more Higgsino-like.  This is known as the `focus point' region\cite{focus}.

\begin{figure}
\centering
	\epsfxsize=10cm
\epsfbox{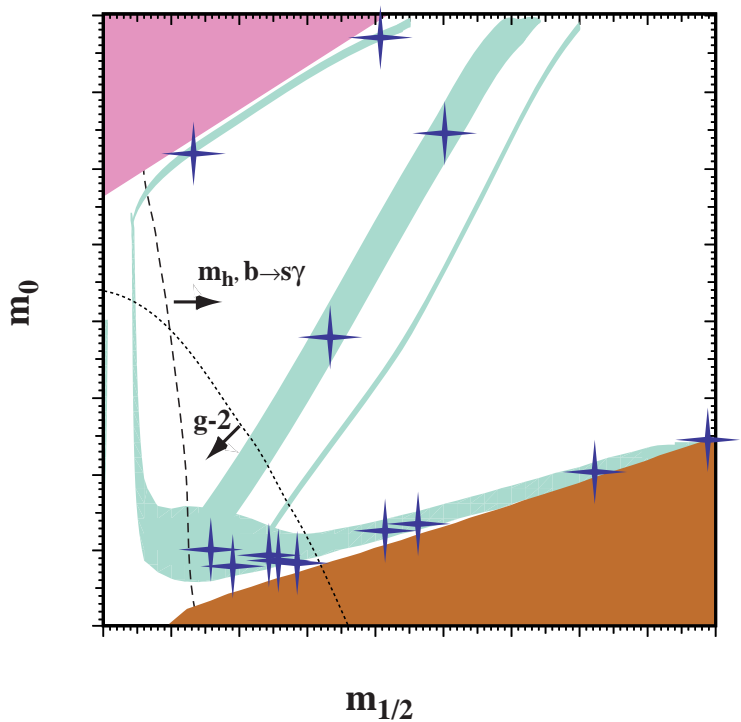}
\caption{Schematic overview of the CMSSM benchmark points proposed 
in~\protect\cite{benchmark}.
 The points are intended to illustrate the range of
available possibilities. The labels correspond to the approximate
positions of the benchmark points in the $(m_{1/2}, m_0)$ plane.
They also span values of $\tan \beta$ from 5  to 50 and include
points with $\mu < 0$. }
\label{fig:Bench}
\end{figure}

As an aid to the assessment of the prospects for detecting sparticles at
different accelerators, benchmark sets of supersymmetric parameters have
often been found useful, since they provide a focus for
concentrated discussion. A set of proposed post-LEP benchmark
scenarios\cite{benchmark} in the CMSSM are illustrated schematically in
Fig.~\ref{fig:Bench}.  Five of the
chosen points are in the
`bulk' region at small $m_{1/2}$ and $m_0$, four are spread along the
coannihilation `tail' at larger $m_{1/2}$ for various values of
$\tan\beta$.  This tail runs along the shaded region in the lower right corner
where the stau is the LSP and is therefore excluded by the constraints against
charged dark matter. 
Two points are in rapid-annihilation `funnels' at large $m_{1/2}$ and $m_0$. 
At large values of $m_0$,  the focus-point region runs along the
boundary where electroweak symmetry no longer occurs (shown in Fig.
\ref{fig:Bench} as the shaded region in the upper left corner). Two points
were chosen in the focus-point region at large $m_0$. The proposed
points range over the allowed values of
$\tan\beta$ between 5 and 50. The light shaded region corresponds to the portion
of parameter space where the relic density $\Omega_\chi h^2$ is
between 0.1 and 0.3. 

\begin{figure}
\begin{center}
\hspace{0.1truecm}
\epsfxsize=9cm   %width of figure - will enlarge/reduce the figures
\epsfbox{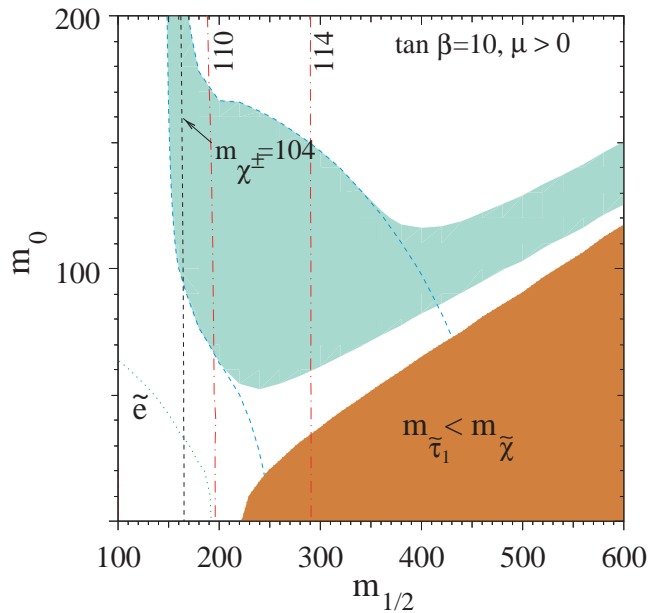}
\caption{The light-shaded `bulk' area is the cosmologically preferred 
region with \protect\mbox{$0.1\leq \Omega h^2 \leq 0.3$}.   The light
dashed lines show the location  of the cosmologically preferred region 
if one ignores  coannihilations with the light sleptons.  
In the dark shaded region in the bottom right, the LSP is
the ${\tilde
\tau}_1$, leading to an unacceptable abundance
of charged dark matter.  Also shown is the isomass
contour $m_{\chi^\pm} = 104$~GeV and $m_h = 110,114$~GeV,
as well as an indication of the slepton bound from
LEP.}
\label{fig:sm}       
\end{center}
\end{figure}

The effect of coannihilations is
to create an allowed band about 25-50 GeV wide in $m_0$ for $m_{1/2} \la
1400$ GeV, which tracks above the $m_{\widetilde \tau}=m_\chi$ contour. 
Along the line $m_{\widetilde \tau}= m_\chi$, $R\approx10$, from
(\ref{eq:R})
\cite{efo}.  As
$m_0$ increases,
the mass difference increases and the slepton contribution to
$\hat\sigma_{\rm  eff}$ falls,  and the relic density rises
abruptly.  This effect is seen in Fig. \ref{fig:sm}.
The light
shaded region corresponds to
\mbox{$0.1<\Omega h^2<0.3$}.  The dark shaded region has $m_{\widetilde
\tau}< m_\chi$ and is excluded. The light dashed contours
indicate the corresponding region in $\Omega h^2$ if one ignores the
effect of coannihilations.  Neglecting coannihilations, one would find an
upper bound of $\sim450$ GeV on $m_{1/2}$, corresponding to an upper bound
of roughly $200$ GeV on $m_{\tilde B}$.  Instead, values of $m_{1/2}$ up to $\sim 1400$ GeV are allowed corresponding to an upper bound
of  $\sim 600$ GeV on $m_{\tilde B}$.

The most important phenomenological constraints are also shown schematically in
Figure~\ref{fig:Bench}.  These include the
constraint provided by the LEP lower
limit on the Higgs mass: $m_H > $ 114.4 GeV \cite{LEPHiggs}.
This holds in the Standard Model, for the lightest Higgs boson
$h$ in the general MSSM for
$\tan\beta
\la 8$, and almost always in the CMSSM for all $\tan\beta$.
Since $m_h$ is sensitive to sparticle masses, particularly
$m_{\tilde t}$, via loop corrections,
the Higgs limit also imposes important constraints on the
CMSSM parameters, principally $m_{1/2}$ as seen by the dashed curve in
Fig.~\ref{fig:Bench}. 

The constraint imposed by
measurements of $b\rightarrow s\gamma$~\cite{bsg} also exclude small values of
$m_{1/2}$. These measurements agree with the Standard Model, and therefore provide
bounds on MSSM particles,  such as the chargino and charged Higgs
masses, in particular. Typically, the $b\rightarrow s\gamma$
constraint is more important for $\mu < 0$, but it is also relevant for
$\mu > 0$,  particularly when $\tan\beta$ is large. 
The BNL E821
experiment reported last year a new measurement of
$a_\mu\equiv {1\over 2} (g_\mu -2)$ which deviated by 2.6 standard
deviations from the best Standard Model prediction available at that
time\cite{BNL}.  However, 
it had been realized that the sign of the most important
pseudoscalar-meson pole part of the light-by-light  scattering
contribution\cite{lightbylight} to the Standard Model prediction should
be reversed, which reduces the apparent experimental discrepancy to about
1.6 standard deviations ($\delta a_\mu \times 10^{10}  = 26
\pm 16$). The largest contribution to the errors in the comparison
with theory was thought to be the statistical error of the experiment,
which has been significantly reduced just recently\cite{BNL2}. 
The world average of $a_\mu\equiv {1\over 2}
(g_\mu -2)$ now deviates by
$(33.9 \pm 11.2) \times 10^{-10}$ from the Standard Model calculation
of Davier et al.\cite{Davier} using $e^+ e^-$ data, and by $(17 \pm 11)
\times 10^{-10}$ from the Standard Model calculation of  Davier et
al.\cite{Davier} based on
$\tau$ decay data. Other recent analyses of the $e^+ e^-$ data yield
similar results. On the subsequent plots,  the formal
2-$\sigma$ range $11.5 \times 10^{-10} < \delta a_\mu < 56.3 \times
10^{-10}$ is displayed.

\begin{figure}[hbtp]
	\centering
		\epsfxsize=10cm
\epsfbox{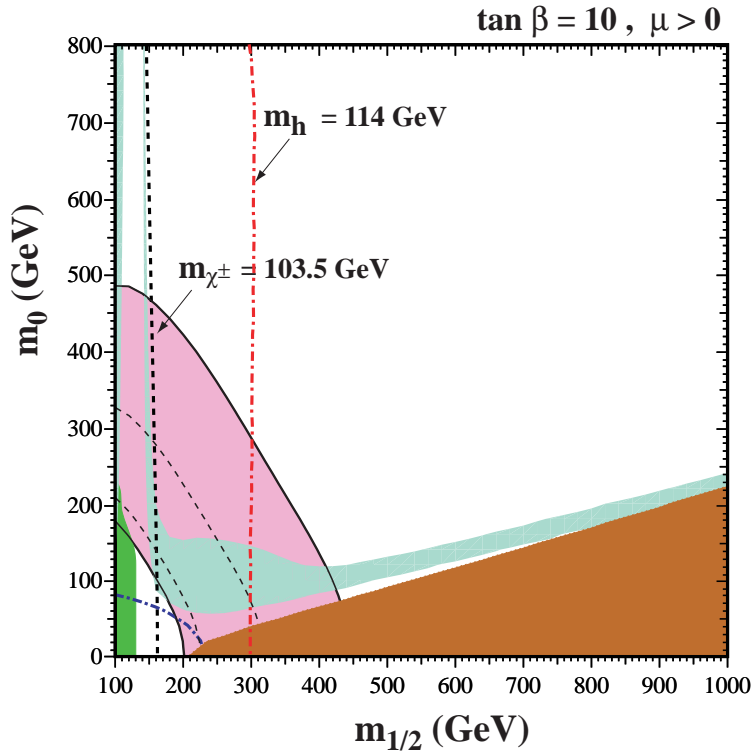}
	\caption{Compilation of phenomenological constraints on the
CMSSM for $\tan \beta = 10,  \mu > 0$,  assuming
$A_0 = 0, m_t = 175$~GeV and $m_b(m_b)^{\overline {MS}}_{SM} =
4.25$~GeV.  The near-vertical lines are the
LEP limits
$m_{\chi^\pm} = 103.5$~GeV (dashed and
black)\protect\cite{LEPsusy}, and
$m_h = 114.1$~GeV (dotted and red)\protect\cite{LEPHiggs}. 
Also, in the lower left corner we show the $m_{\tilde e}
= 99$ GeV contour\protect\cite{LEPSUSYWG_0101}.  In the dark
(brick red) shaded regions, the LSP is the charged
${\tilde \tau}_1$, so this region is excluded. The
light(turquoise) shaded areas are the cosmologically preferred
regions with
\protect\mbox{$0.1\leq \Omega h^2 \leq
0.3$}~\protect\cite{EFGOSi}. The
medium (dark green) shaded regions  are excluded by $b \to s
\gamma$~\protect\cite{bsg}. The shaded (pink) region in the 
upper right delineates the
$2 \, \sigma$ range of $g_\mu - 2$.  The dashed curves within this region correspond to the $1-\sigma$ bounds. }
	\label{rd2c}
\end{figure}

Following a previous analysis\cite{EFGOSi,eos2}, in Figure
\ref{rd2c} the $m_{1/2}-m_0$ parameter space is shown for $\tan
\beta = 10$.   The dark shaded region (in the lower right)
corresponds to the parameters where the LSP is not a neutralino
but rather a
${\tilde \tau}_R$. 
 The cosmologically interesting region at
the left of the figure is due to the appearance
of pole effects.  There, the LSP can annihilate through
s-channel $Z$ and $h$ (the light Higgs) exchange, thereby
allowing a very large value of $m_0$. However, this region is
excluded by phenomenological constraints. Here one can see clearly
the coannihilation tail which extends towards large values of
$m_{1/2}$. In addition to the phenomenological constraints discussed above,
Figure
\ref{rd2c} also shows the current experimental constraints on
the CMSSM parameter space due to the limit $m_{\chi^\pm}
\ga$ 103.5 GeV provided by chargino searches at  LEP
\cite{LEPsusy}. LEP has also provided lower limits on
slepton masses, of which the strongest is $m_{\tilde e}\ga$ 99
GeV \cite{LEPSUSYWG_0101}. This is shown by dot-dashed curve in the
lower left corner of Fig. \ref{rd2c}. Similar results have been found by other
analyses\cite{otherOmega}.

As one can see, one of the most important phenomenological constraint at this value of
$\tan \beta$ is due to the Higgs mass (shown by the nearly vertical dot-dashed
curve).  The theoretical Higgs masses were evaluated using {\tt
FeynHiggs}\cite{FeynHiggs}, which is estimated to  have a residual uncertainty of a
couple of GeV in $m_h$. The region excluded
by the  $b\rightarrow s\gamma$ constraint is the dark shaded (green)
region to the left of the plot. As many authors have pointed
out\cite{susygmu}, a discrepancy between theory and the BNL experiment
could well be explained by supersymmetry. As seen in Fig.~\ref{rd2c},
this is particularly easy if $\mu > 0$. The
medium (pink) shaded region in the figure corresponds to the overall
allowed region ($2\sigma$) by the new experimental result.

\begin{figure}[hbtp]
	\centering 
		\epsfxsize=10.0cm
\epsfbox{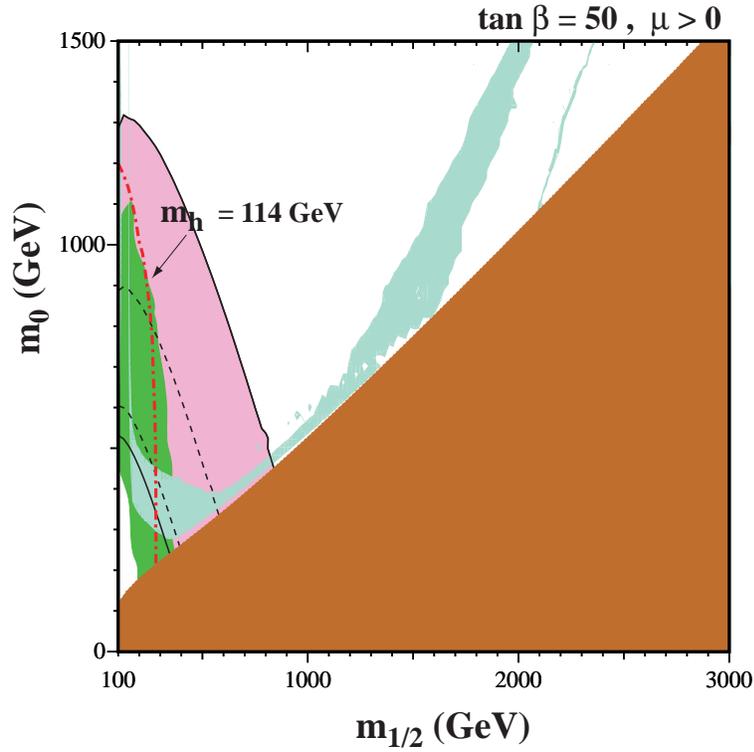}
	\caption{As in Fig. \protect\ref{rd2c} for $\tan \beta = 50$. }
	\label{rd2c50}
\end{figure}

As discussed above, another
mechanism for extending the allowed CMSSM region to large
$m_\chi$ is rapid annihilation via a direct-channel pole when $m_\chi
\sim {1\over 2} m_{A}$~\cite{funnel,EFGOSi}. This may yield a
`funnel' extending to large $m_{1/2}$ and $m_0$ at large $\tan\beta$, as
seen in Fig.~\ref{rd2c50}.

In principle, the true 
input parameters in the CMSSM are: $\mu, m_1, m_2,$ and $ B$, where $m_1$ and $m_2$
are the Higgs soft masses (in the CMSSM $m_1 = m_2 = m_0$ and $B$ is the
susy breaking bilinear mass term). In this case, the electroweak symmetry breaking
conditions lead to	a prediction of $M_Z, \tan \beta$ ,and  $m_A$.  Since we are 
not really interested in predicting $M_Z$, it is more useful to assume  
instead the following CMSSM	input parameters: $M_Z, m_1, m_2,$ and $\tan \beta$
again with	$m_1 = m_2 = m_0$. In this case, one predicts $\mu, B$, and $m_A$.
However, one can generalize the CMSSM case to include non-universal Higgs masses\cite{nonu,eos3}
(NUHM), in which case the 			
input parameters become:$ M_Z, \mu, m_A,$ and $\tan \beta$ and one predicts $m_1,
m_2$, and $B$. 

\begin{figure}[hbtp]
	\centering 
		\epsfxsize=9.5cm
\epsfbox{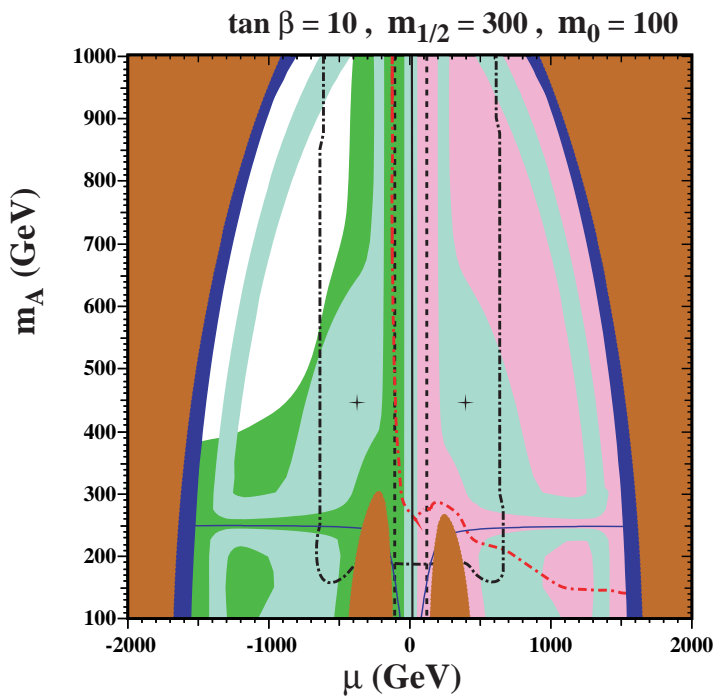}
	\caption{Compilations of phenomenological constraints on the MSSM with NUHM 
in the $(\mu, m_A)$ plane for $\tan \beta = 10$ and $m_0 = 100$~GeV, 
$m_{1/2} = 300$~GeV, assuming $A_0 = 0$, $m_t = 175$~GeV and 
$m_b(m_b)^{\overline {MS}}_{SM} = 4.25$~GeV.
The  shading is as described in Fig.~\protect\ref{rd2c}.
The (blue) solid line is the contour $m_\chi = m_A/2$, near which
rapid direct-channel annihilation suppresses the relic density. 
The dark (black) dot-dashed line indicates when one or another Higgs 
mass-squared becomes negative at the GUT scale: only lower $|\mu|$ and 
larger $m_A$ values are allowed. The crosses denote the values of 
$\mu$ and $m_A$ found in the CMSSM.}
	\label{muma}
\end{figure}

The NUHM parameter space was recently analyzed\cite{eos3} and a sample of the results
found is shown in Fig. \ref{muma}. While much of the cosmologically preferred area
with $\mu < 0$ is excluded, there is a significant enhancement in the allowed 
parameter space for $\mu > 0$. 

\subsection{Detection}

Because the LSP as dark matter is present locally, there are many
avenues for pursuing dark matter detection. Direct detection techniques
rely on an ample neutralino-nucleon scattering cross-section.
The effective four-fermion lagrangian can be written as
\begin{eqnarray}
\mathcal{L}  & =  &\bar{\chi} \gamma^\mu \gamma^5 \chi \bar{q_{i}} 
\gamma_{\mu} (\alpha_{1i} + \alpha_{2i} \gamma^{5}) q_{i}  \nonumber \\
& + & \alpha_{3i} \bar{\chi} \chi \bar{q_{i}} q_{i} + 
\alpha_{4i} \bar{\chi} \gamma^{5} \chi \bar{q_{i}} \gamma^{5} q_{i} \nonumber \\
& + &\alpha_{5i} \bar{\chi} \chi \bar{q_{i}} \gamma^{5} q_{i} +
\alpha_{6i} \bar{\chi} \gamma^{5} \chi \bar{q_{i}} q_{i} 
\end{eqnarray}
However, the terms involving $\alpha_{1i}, \alpha_{4i}, \alpha_{5i}
$, and
$\alpha_{6i}$  lead to velocity dependent  elastic cross sections.
The remaining terms are: the spin dependent coefficient,
$\alpha_{2i} $  and the scalar coefficient $\alpha_{3i} $.
Contributions to $\alpha_{2i} $ are predominantly through light squark exchange. 	
This is the dominant channel for binos.
Scattering also occurs through Z exchange but this channel 
requires a strong Higgsino component.
Contributions to $\alpha_{3i} $  are also dominated by
light squark exchange 	
but Higgs exchange
is non-negligible in most cases. 

\begin{figure}
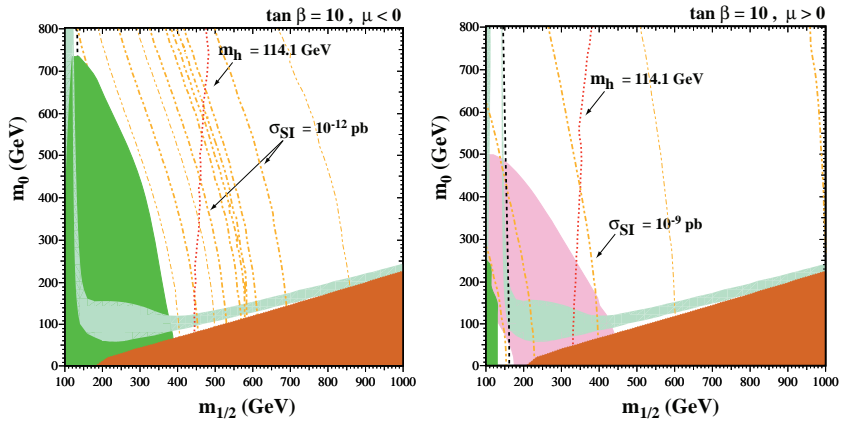

%\vspace*{-0.75in}
%\hspace*{-.70in}
		\epsfxsize=5.5cm
\epsfbox{si10n.epss}
%\vspace*{-6.65in}
%\hspace*{2.2in}
\epsfxsize=5.5cm
\epsfbox{si10p.epss} 
%\vskip 2.5in
\caption{\label{fig:sicontours}
{\it Spin-independent cross sections in the $(m_{1/2}, m_0)$ planes for
(a) $\tan \beta = 10, \mu < 0$,  (b) $\tan \beta = 10, \mu > 0$.   The
double dot-dashed (orange) curves are contours of the spin-independent
cross section, differing by factors of 10 (bolder) and interpolating
factors of 3 (finer - when shown). For example, in (b), the curves to the
right of the one marked $10^{-9}$ pb correspond to $3
\times  10^{-10}$~pb and $10^{-10}$~pb. 
}}
\end{figure}

Fig.~\ref{fig:sicontours} displays contours of the
spin-independent cross section for the elastic scattering of the LSP
$\chi$ on protons in the $m_{1/2}, m_0$ planes for (a) $\tan \beta = 10,
\mu < 0$, (b) $\tan \beta = 10, \mu > 0$ \cite{eflo3}. The double
dot-dashed (orange) lines are contours of the spin-independent cross
section, and the contours
$\sigma_{SI} = 10^{-9}$~pb in panel (a) and $\sigma_{SI} =
10^{-12}$~pb in panel (b) are
indicated.
The LEP lower limits on $m_h$ and $m_{\chi^\pm}$, as well as the
experimental measurement of $b \to s \gamma$ for $\mu < 0$, tend to bound
the cross sections from above, as discussed in more
detail below. Generally speaking, the spin-independent cross section is 
relatively large in the `bulk' region, but falls off in the coannihilation
`tail'.   Also, we note also that there is a strong
cancellation in the spin-independent cross section when
$\mu < 0$~\cite{EFlO1,EFlO2}, as seen along strips in panel (a)  of
Fig.~\ref{fig:sicontours} where
$m_{1/2} \sim 500$~GeV. In the cancellation region,
the cross section drops lower than $10^{-14}$ pb. All these possibilities
for suppressed spin-independent cross sections are disfavoured by the
data on
$g_\mu - 2$, which favour values of $m_{1/2}$ and
$m_0$ that are not very large, as well as $\mu > 0$, as seen in panel
(b) of Fig.~\ref{fig:sicontours}. Thus $g_\mu - 2$ tends to provide a
lower bound on the spin-independent cross section.

\begin{figure}
%\vspace*{-0.75in}
\begin{center}
		\epsfysize=4.8 cm
\epsfbox{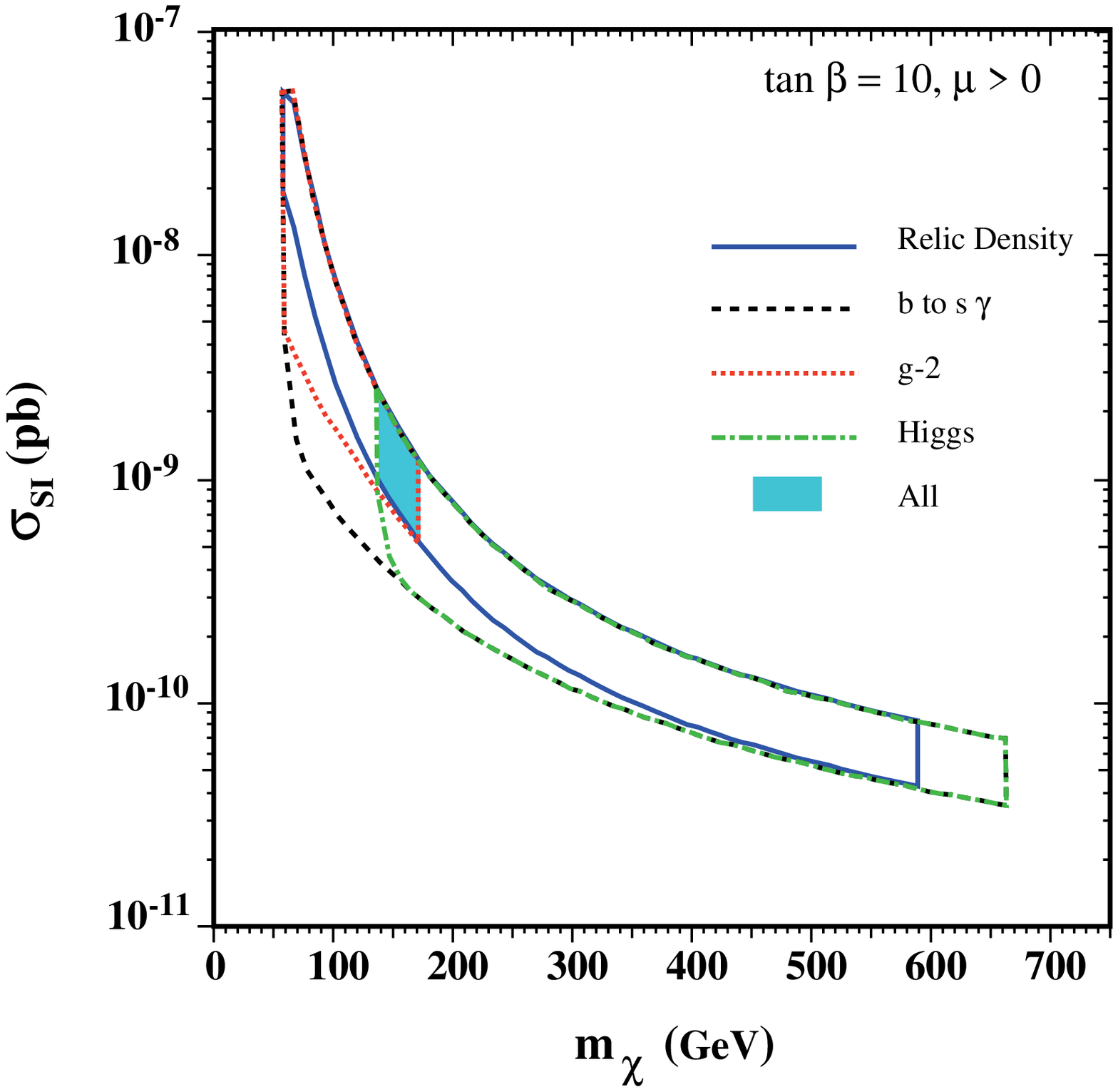}
\epsfysize=4.8 cm
\epsfbox{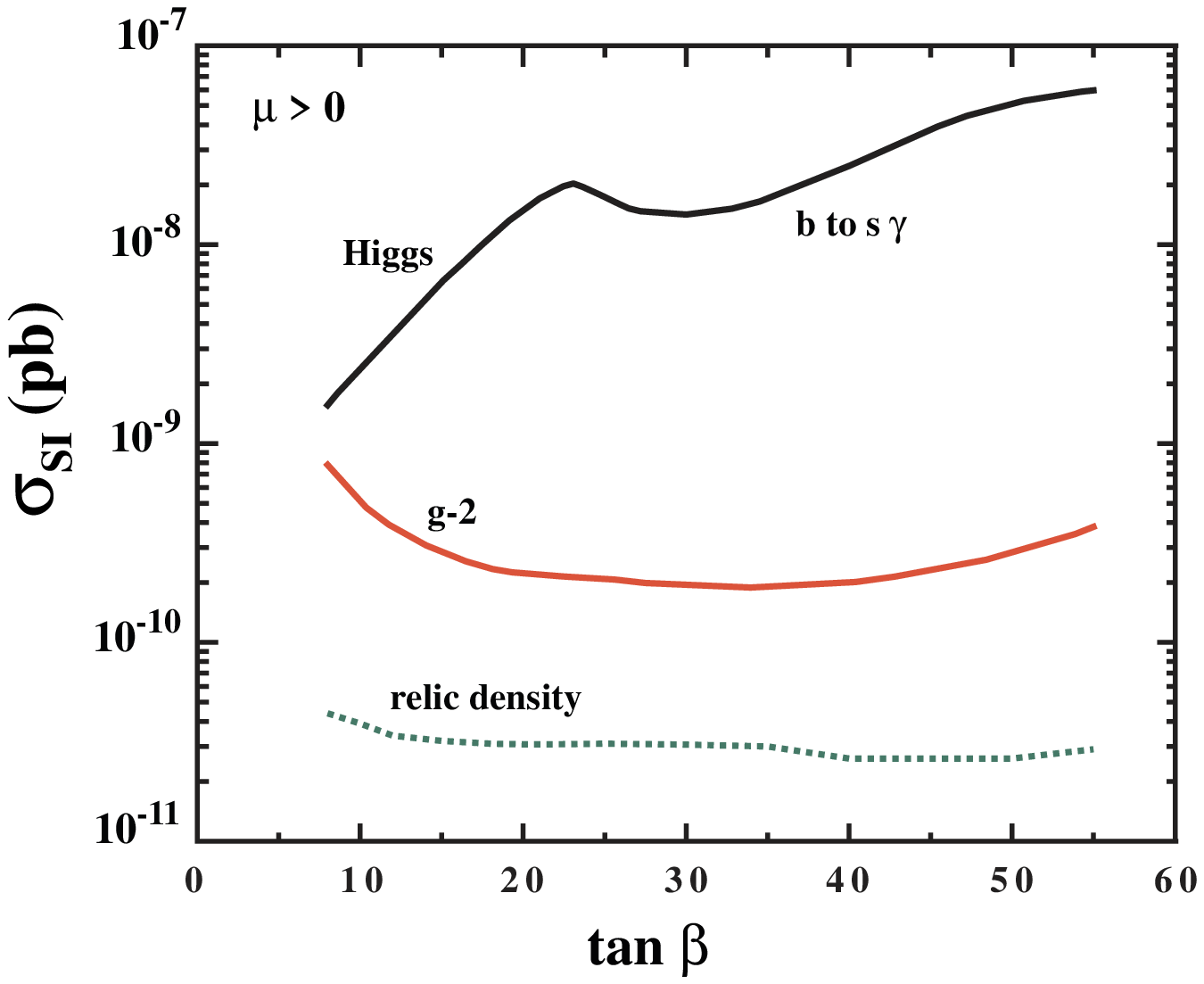}
\end{center}
%\vspace*{-3in}
\caption{\label{fig:decimation}
{\it Allowed ranges of the cross sections for $\tan \beta = 10$ (a)
$\mu > 0$ for spin-independent elastic scattering.  The solid (blue)
lines  indicate the
relic density constraint, the dashed (black) lines the $b
\to s \gamma$ constraint, the dot-dashed (green) lines the
$m_h$ constraint, and the dotted (red) lines the $g_\mu -
2$ constraint. The shaded (pale blue)  region is allowed by all
the constraints. (b) The allowed ranges of the spin-independent cross
section for  $\mu > 0$. The darker solid (black) lines show the upper 
limits on 
the cross sections obtained from $m_h$ and $b \to s \gamma$, and (where 
applicable) the lighter 
solid (red) lines show the lower limits suggested by $g_\mu - 2$ and the 
dotted (green) lines the lower limits from the relic density.
}}
\end{figure}

Fig.~\ref{fig:decimation}(a) illustrates the effect on the cross sections
of each of the principal phenomenological constraints, for the particular
case $\tan \beta = 10$  $\mu > 0$. The solid
(blue) lines mark the bounds on the cross sections allowed by the
relic-density constraint $0.1 < \Omega_\chi h^2 < 0.3$
alone. 
For any given value of
$m_{1/2}$, only a restricted range of $m_0$ is allowed. Therefore, only a
limited range of $m_0$, and hence only a limited range for the cross
section, is allowed for any given value of
$m_\chi$. The thicknesses of the allowed regions are due in part to the 
assumed
uncertainties in the nuclear inputs.  These have been discussed at
length in \cite{EFlO2,EFlO1}.
 On 
the other hand, a broad range of $m_\chi$ is allowed, when one takes into
account the coannihilation `tail' region at each $\tan 
\beta$ and the rapid-annihilation `funnel' regions for $\tan
\beta = 35, 50$. The dashed (black) line  displays the range allowed by
the $b \to s
\gamma$ constraint alone. In
this case, a broader range of $m_0$ and hence the spin-independent cross
section is possible for any given value of $m_\chi$. The impact of the
constraint due to $m_h$ is shown by the dot-dashed (green) line. 
Comparing with the previous constraints, we see that a region at low
$m_\chi$ is excluded by $m_h$, strengthening significantly the previous
{\it upper} limit on the spin-independent cross section. Finally, the
dotted (red)  lines in Fig.~\ref{fig:decimation} show the impact of the
$g_\mu - 2$ constraint. This imposes an upper bound on
$m_{1/2}$ and hence $m_\chi$, and correspondingly a {\it lower} limit on
the spin-independent cross section.

\begin{figure}[hbtp]
	\centering 
		\epsfxsize=9.5cm
\epsfbox{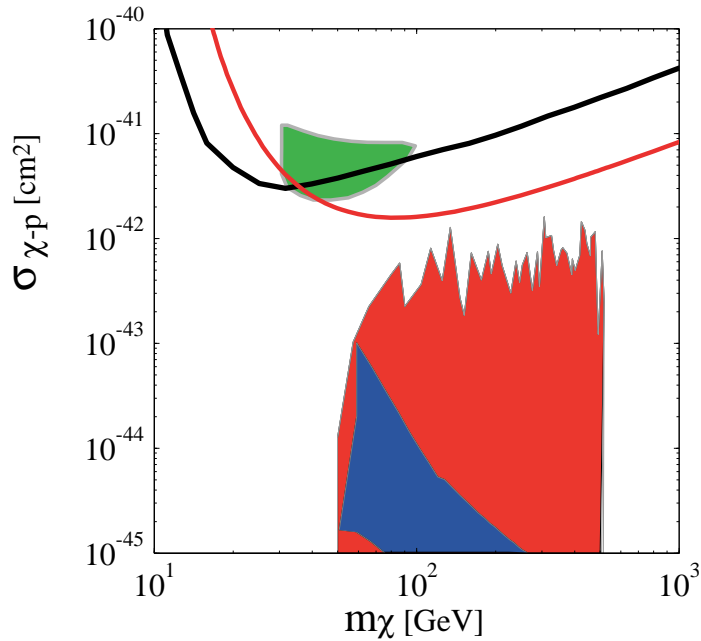}
	\caption{Limits from the CDMS\protect\cite{cdms} and Edelweiss\protect\cite{edel}
experiments on the neutralino-proton elastic scattering cross section as a function
of the neutralino mass. The Edelweiss limit is stronger at higher $m_\chi$. These
results nearly exclude the shaded region observed by DAMA\protect\cite{dama}. The
theoretical predictions lie at lower values of the cross section.}
	\label{cdms}
\end{figure}

This analysis is 
extended 
in panel (b)  of Fig.~\ref{fig:decimation} to all the values $8 
< \tan \beta \le 55$ and we find overall that \cite{eflo3}
\begin{eqnarray}
2 \times 10^{-10}~{\rm pb} \la & \sigma_{SI} & \la 6 \times 10^{-8}~{\rm
pb},
\\ 2 \times 10^{-7}~{\rm pb} \la & \sigma_{SD} & \la  10^{-5}~{\rm
pb},
\label{xsecrangeallp}
\end{eqnarray}
for $\mu > 0$. ($\sigma_{SD}$ is the spin-dependent cross-section not
shown in the figures presented here.) As we see in panel (b)  of
Fig.~\ref{fig:decimation},
$m_h$ provides the most  important upper limit 
on the cross sections for $\tan \beta < 23$, and $b \to s \gamma$ for 
larger $\tan \beta$, with $g_\mu - 2$ always providing a more stringent 
lower limit than the relic-density constraint. 
The relic density constraint shown is evaluated at the endpoint of the
coannihilation region. At large $\tan \beta$,  the
Higgs funnels or the focus-point regions have not been considered, as
their locations are very  sensitive to input parameters and calculational
details\cite{EO}.

The results from a CMSSM and MSSM analysis\cite{EFlO1,EFlO2} for $\tan \beta = 3$ and
10 are compared with the most recent CDMS\cite{cdms} and Edelweiss\cite{edel}
bounds in Fig.~\ref{cdms}. These results have nearly entirely excluded the
region purported by the DAMA\cite{dama} experiment. The CMSSM prediction\cite{EFlO1} is shown  by the dark shaded region, while the NUHM case\cite{EFlO2} is
shown by the larger lighter shaded region. Other CMSSM results\cite{otherCMSSMDM}
are also available.

\begin{figure}
\centering
		\epsfxsize=10cm
\epsfbox{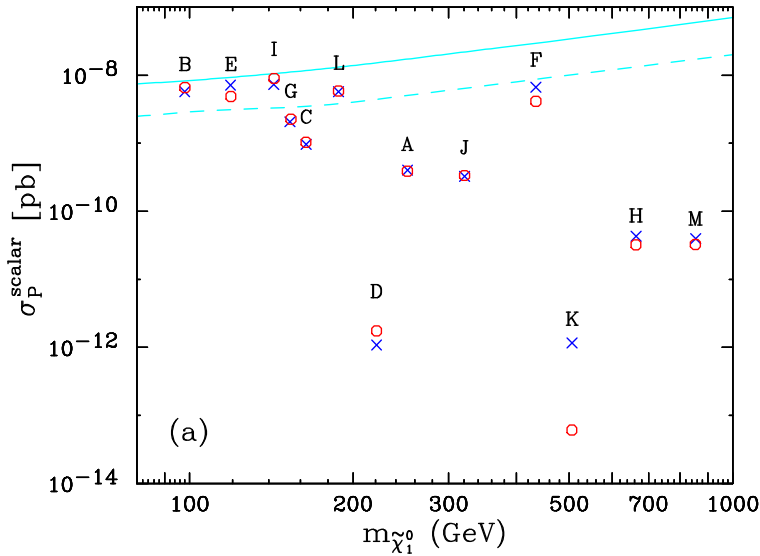}
\caption{ Elastic spin-independent scattering  
of supersymmetric relics on protons  calculated in 
benchmark scenarios\protect\cite{EFFMO}, compared with the 
projected sensitivities for CDMS
II~\protect\cite{Schnee:1998gf} and CRESST\protect\cite{Bravin:1999fc}
(solid) and GENIUS\protect\cite{GENIUS} (dashed).
The predictions of our code (blue
crosses) and {\tt Neutdriver}\protect\cite{neut} (red circles) for
neutralino-nucleon scattering are compared.
The labels A, B, ...,L correspond to the benchmark points as shown in 
Fig.~\protect\ref{fig:Bench}.}
\label{fig:DM}
\end{figure}

I conclude by showing the
prospects for direct detection for the benchmark points discussed
above\cite{EFFMO}. Fig.~\ref{fig:DM} shows rates for the elastic
spin-independent scattering of supersymmetric relics,
including the projected sensitivities for CDMS
II\cite{Schnee:1998gf} and CRESST\cite{Bravin:1999fc} (solid) and
GENIUS\cite{GENIUS} (dashed).
Also shown are the cross sections 
calculated in the proposed benchmark scenarios discussed in the previous
section, which are considerably below the DAMA\cite{dama} range
($10^{-5} - 10^{-6}$~pb).
Indirect searches for supersymmetric dark matter via the products of
annihilations in the galactic halo or inside the Sun also have prospects
in some of the benchmark scenarios\cite{EFFMO}.

\end{document}